\documentclass[aps,prb,reprint,superscriptaddress]{revtex4-2}

\usepackage{amsmath}
\usepackage{upgreek}
\usepackage{float}
\usepackage{graphicx}
\usepackage{tabularx}
\usepackage{hyperref}

\begin{document}

\title{Closing the gap between theory and experiment for lithium manganese oxide spinels using a high-dimensional neural network potential}

\author{Marco Eckhoff}
\email{marco.eckhoff@chemie.uni-goettingen.de}
\affiliation{Universit\"at G\"ottingen, Institut f\"ur Physikalische Chemie, Theoretische Chemie, Tammannstra{\ss}e 6, 37077 G\"ottingen, Germany.}
\author{Florian Sch\"onewald}
\affiliation{Universit\"at G\"ottingen, Institut f\"ur Materialphysik, Friedrich-Hund-Platz 1, 37077 G\"ottingen, Germany.}
\author{Marcel Risch}
\affiliation{Universit\"at G\"ottingen, Institut f\"ur Materialphysik, Friedrich-Hund-Platz 1, 37077 G\"ottingen, Germany.}
\affiliation{Helmholtz-Zentrum Berlin für Materialien und Energie GmbH, Nachwuchsgruppe Gestaltung des Sauerstoffentwicklungsmechanismus, Hahn-Meitner-Platz 1, 14109 Berlin, Germany.}
\author{Cynthia A. Volkert}
\affiliation{Universit\"at G\"ottingen, Institut f\"ur Materialphysik, Friedrich-Hund-Platz 1, 37077 G\"ottingen, Germany.}
\affiliation{Universit\"at G\"ottingen, International Center for Advanced Studies of Energy Conversion (ICASEC), Tammannstra{\ss}e 6, 37077 G\"ottingen, Germany.}
\author{Peter E. Bl\"ochl}
\affiliation{Technische Universit\"at Clausthal, Institut f\"ur Theoretische Physik, Leibnizstra{\ss}e 10, 38678 Clausthal-Zellerfeld, Germany.}
\affiliation{Universit\"at G\"ottingen, Institut f\"ur Theoretische Physik, Friedrich-Hund-Platz 1, 37077 G\"ottingen, Germany.}
\author{J\"org Behler}
\email{joerg.behler@uni-goettingen.de}
\affiliation{Universit\"at G\"ottingen, Institut f\"ur Physikalische Chemie, Theoretische Chemie, Tammannstra{\ss}e 6, 37077 G\"ottingen, Germany.}
\affiliation{Universit\"at G\"ottingen, International Center for Advanced Studies of Energy Conversion (ICASEC), Tammannstra{\ss}e 6, 37077 G\"ottingen, Germany.}

\date{\today}

\begin{abstract}
Many positive electrode materials in lithium ion batteries include transition metals which are difficult to describe by electronic structure methods like density functional theory (DFT) due to the presence of multiple oxidation states. A prominent example is the lithium manganese oxide spinel Li$_x$Mn$_2$O$_4$ with $0\leq x\leq2$. While DFT, employing the local hybrid functional PBE0r, provides a reliable description, the need for extended computer simulations of large structural models remains a significant challenge. Here, we close this gap by constructing a DFT-based high-dimensional neural network potential (HDNNP) providing accurate energies and forces at a fraction of the computational costs. As different oxidation states and the resulting Jahn-Teller distortions represent a new level of complexity for HDNNPs, the potential is carefully validated by performing X-ray diffraction experiments. We demonstrate that the HDNNP provides atomic level details and is able to predict a series of properties like the lattice parameters and expansion with increasing Li content or temperature, the orthorhombic to cubic transition, the lithium diffusion barrier, and the phonon frequencies. We show that for understanding these properties access to large time and length scales as enabled by the HDNNP is essential to close the gap between theory and experiment. 
\end{abstract}


\maketitle

\section{Introduction}

Advances in battery technology are more important than ever to provide a reliable energy supply for countless applications, from portable electronic devices to electric vehicles, and in particular lithium ion batteries have gained a central role \cite{Tarascon2001, Goodenough2010}. One important example for positive electrode materials in lithium ion batteries is the lithium manganese oxide spinel Li$_x$Mn$_2$O$_4$ with $0\leq x\leq2$ \cite{Thackeray1983, Thackeray1997, Nitta2015}, which is the subject of this study. Due to its abundance and non-toxicity Li$_x$Mn$_2$O$_4$ is more environmentally friendly than other lithium ion positive electrode materials \cite{Tarascon2001}. Furthermore, it has also recently been used as an electrocatalyst \cite{Cady2015, Koehler2017, Baumung2019a}.

\begin{figure}[tb!]
\centering
\includegraphics[width=\columnwidth]{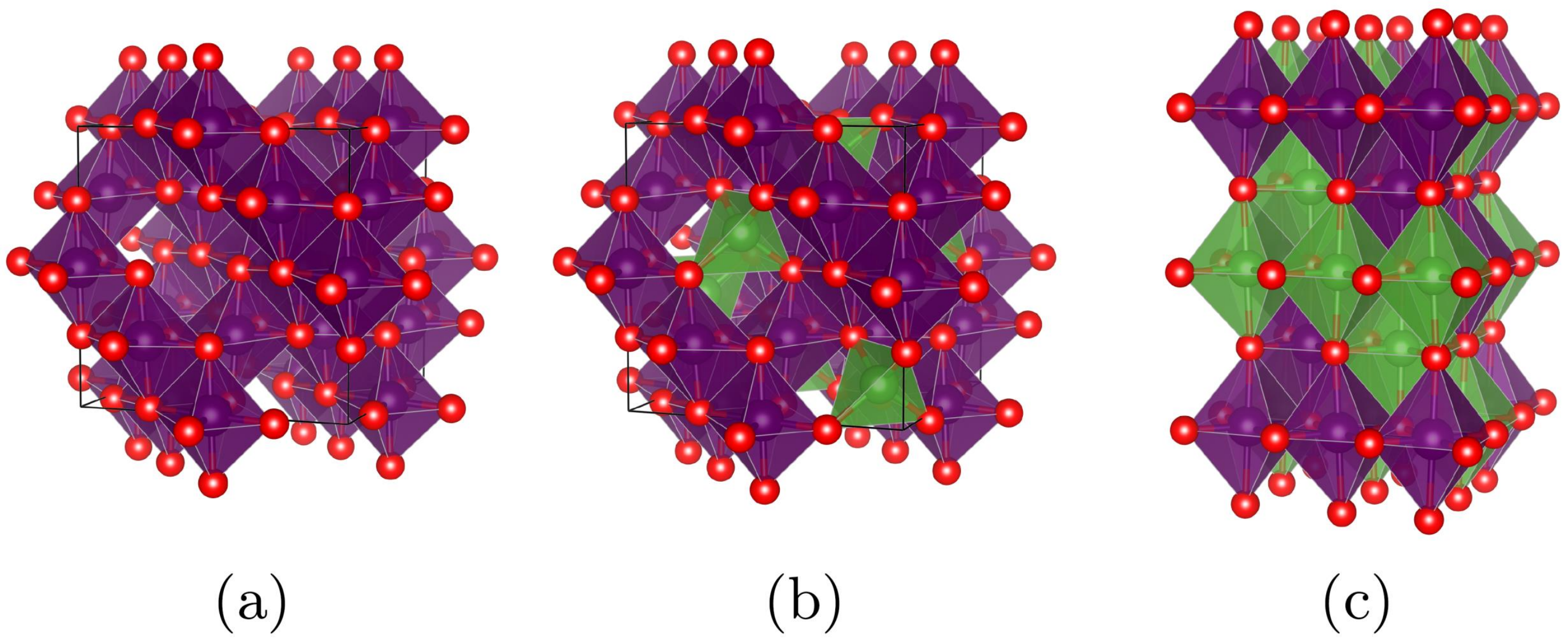}
\caption{Li$_x$Mn$_2$O$_4$ spinel structure with (a) $x=0$ \cite{Takahashi2003}, (b) $x=1$ \cite{Akimoto2004}, and (c) $x=2$ \cite{Mosbah1983}. Li atoms and their coordination polyhedra are shown in green, O atoms in red, and Mn atoms and their coordination polyhedra in purple. The black lines represent the unit cell. The figure was created with VESTA version 3.4.4 \cite{Momma2011}.}\label{fig_limn2o4_structure}
\end{figure}

The crystal structure of Li$_x$Mn$_2$O$_4$ is a cubic spinel with space group Fd$\overline{3}$m in the range from $0\leq x\leq1$ and above $\sim$\,290\,K (Figures \ref{fig_limn2o4_structure} (a) and (b)) \cite{Takahashi2003, Akimoto2004}. The $\uplambda$-Mn$_2$O$_4$ ($x=0$) structure is based on MnO$_6$ octahedra. These share half of their edges and form a superstructure of $\left(\text{MnO}_6\right)_4$ tetrahedra, which themselves are corner-sharing. Using electrochemical intercalation, for example, Li can be inserted in tetrahedral sites \cite{Tarascon1994}. In LiMn$_2$O$_4$ every corner of a MnO$_6$ octahedron is shared with one LiO$_4$ tetrahedron. The Li positions build two individual fcc sublattices, i.e., a diamond structure. In the range from $x=0$ to $1$ the Li intercalation leads to an almost linearly increasing lattice constant without major changes of the manganese oxide host lattice \cite{Thackeray1984,Ohzuku1989}. For $x>1$ a tetragonal spinel forms with $x=2$ and space group I4$_1$/amd (Figure \ref{fig_limn2o4_structure} (c)) \cite{Mosbah1983}, which coexists with the $x=1$ cubic spinel in the range $1<x<2$ \cite{Ohzuku1989, Maier2016}. In Li$_2$Mn$_2$O$_4$, the Li ions occupy octahedral sites.

Variations in the Li content change the oxidation states of the Mn ions \cite{Berg1999}. While in $\uplambda$-Mn$_2$O$_4$ all Mn ions are in the Mn$^\text{IV}$ state, Li$_2$Mn$_2$O$_4$ exclusively contains Mn$^\text{III}$ ions. In stoichiometric LiMn$_2$O$_4$ the numbers of Mn$^\text{III}$ and Mn$^\text{IV}$ ions are equal. The Mn$^\text{III}$ ions are in the high-spin (hs) configuration with the d orbital occupation t$_\mathrm{2g}^3$e$_\mathrm{g}^1$ leading to Jahn-Teller distorted Mn$^\text{III}$O$_6$ octahedra \cite{Jahn1937}. This is the reason for the formation of the tetragonal spinel structure for $x=2$. The cubic spinel structure present for $0<x\leq1$ is a result of the disordered Mn$^\text{III}$/Mn$^\text{IV}$ arrangement \cite{Ouyang2009} in combination with the thermal averaging of the spatial orientations of the Jahn-Teller distorted octahedra \cite{Piszora2004}. Below\ $\sim$\,290\,K the order increases yielding an orthorhombic spinel with space group Fddd for LiMn$_2$O$_4$ \cite{Akimoto2004}.

An accurate description of the narrow, partially filled Mn d bands in Li$_x$Mn$_2$O$_4$ is challenging for electronic structure methods like density functional theory (DFT), and both the local density approximation as well as the generalized gradient approximation (GGA) are not able to provide the correct electronic structure of Li$_x$Mn$_2$O$_4$ \cite{Mishra1999, vanderVen2000, Grechnev2002, Ouyang2009}. Therefore, at least the level of GGA$+U$ or hybrid functionals is required to obtain separate Mn$^\mathrm{IV}$ (t$_\mathrm{2g}^3$e$_\mathrm{g}^0$) and Jahn-Teller distorted hs-Mn$^\mathrm{III}$ (t$_\mathrm{2g}^3$e$_\mathrm{g}^1$) ions as well as a finite band gap \cite{Ouyang2009, Chevrier2010}. In a previous study \cite{Eckhoff2020} we investigated in detail which Hartree-Fock exchange terms have to be added to GGA functionals in order to obtain the correct electronic structure with minimal computational effort. Our extensive benchmark of various exchange-correlation functionals for Li$_x$Mn$_2$O$_4$ and numerous related materials \cite{Eckhoff2020} showed that the local hybrid functional PBE0r \cite{Sotoudeh2017}, which relies on on-site Hartree-Fock exchange only, enables to combine the accuracy of conventional hybrid functionals with the efficiency of GGA calculations.

Specifically, our previous work \cite{Eckhoff2020} focused on the electronic structure of bulk Li$_x$Mn$_2$O$_4$ unit cells at 0\,K. Despite the good agreement with experimental data, DFT does not enable studies of substantially larger, more realistic model systems at finite temperatures due to its high computational effort. However, dynamic studies on nanosecond timescales are required, for example, to investigate the transition from the orthorhombic to the cubic crystal structure at about 290\,K \cite{Akimoto2004} or to examine the (de)lithiation process with associated long-distance diffusion, structural changes, and a very inhomogeneous Li distribution. A more efficient atomistic potential, which allows to avoid the demanding electronic structure calculations and provides the energies and forces directly as a function of the atomic positions, is required for this purpose.

Machine learning has been demonstrated to be a promising tool to develop highly accurate and efficient atomistic potentials \cite{Behler2011a, Behler2016, Botu2017, Deringer2019}. Machine learning potentials are constructed using a large number of flexible functions, which are adapted to reference data from electronic structure calculations to represent the potential energy surface. Potentials based on neural networks \cite{Blank1995, Behler2007a, Behler2007, Faraji2017, Smith2017, Schuett2018}, Gaussian approximation potentials \cite{Bartok2010,Bartok2015}, and many others \cite{Thompson2015, Shapeev2016} have been applied very successfully to describe various types of systems like molecules in the gas phase \cite{Li2013, Gastegger2015, Lu2016, Gastegger2017}, liquids and solutions \cite{Morawietz2016, Cheng2016, Hellstrom2016}, solid/liquid interfaces \cite{Natarajan2016, Quaranta2017}, and solid bulk materials \cite{Behler2008, Eckhoff2019} as well as surface supported clusters \cite{Artrith2013}. However up until now, machine learning potentials have not been applied to systems like Li$_x$Mn$_2$O$_4$ with a complex electronic structure including different oxidation states depending on the Li content, which may or may not exhibit Jahn-Teller distortions.

In this work we develop and validate a high-dimensional neural network potential (HDNNP), a type of machine learning potential suitable for very large systems that has been suggested by Behler and Parrinello \cite{Behler2007, Behler2011, Behler2017}, to represent the potential energy surface of bulk Li$_x$Mn$_2$O$_4$. HDNNPs can autonomously identify, for example in molecular dynamics (MD) simulations, if a prediction is reliable or if a newly visited configuration has to be added to the training data set \cite{Artrith2012,Behler2015}.
The reference data set consists of potential energies and atomic forces of small periodic Li$_x$Mn$_2$O$_4$ cells obtained from the PBE0r functional \cite{Sotoudeh2017, Eckhoff2020} including D3 dispersion corrections \cite{Grimme2010, Grimme2011}. We performed DFT calculations for a wide range of structural motifs including thermally distorted configurations.

Since Li$_x$Mn$_2$O$_4$ is used as positive electrode material, Li diffusion barriers as well as structural deformations during charging and discharging are important physical properties. These are studied as well as structural changes depending on composition and temperature, most prominently the transition from the orthorhombic to the cubic structure occurring near room temperature. In order to validate the HDNNP, we compare to DFT data as well as to in situ X-ray diffraction (XRD) experiments where we investigate the evolution of the crystal structure of Li$_x$Mn$_2$O$_4$ nanoparticles while varying the Li content and the temperature. 

This paper is organized as follows: Firstly, we outline the key concepts of the PBE0r functional and the HDNNP method and present a modified initialization algorithm for the weights in the atomic neural networks of the HDNNP. After a summary of the computational and experimental details, which includes a description how we obtained a reliable and consistent reference data set for a material with a complex electronic structure, we discuss the accuracy of our HDNNP. The validation is carried out using DFT and experimental data including diffusion barriers, electrochemical potentials, and vibrational fingerprints. Subsequently, the HDNNP is applied to study the influence of Li content and temperature variations on the structure of Li$_x$Mn$_2$O$_4$. We reveal the required simulation time and lengths scales to match experimental measurements in the investigation of the orthorhombic to cubic transition. We find that ab initio molecular dynamics would not be applicable to study the transition due to the large computational effort, but that the HDNNP is sufficiently efficient and accurate to yield excellent agreement with experimental data. Employing the HDNNP we are able to characterize the underlying atomistic processes, which give rise to the macroscopic phenomena observed in experiment. 

\section{Theoretical methods}

\subsection{PBE0r local hybrid density functional}

DFT is in principle exact for the energetics of the electronic ground state \cite{Hohenberg1964, Kohn1965}. However, since the explicit form of the exchange-correlation functional is unknown, numerous functionals have been proposed providing different levels of accuracy \cite{Burke2012, Neugebauer2013, Peverati2014}. Hybrid functionals, which are based on GGA functionals mixed with some amount of exact Hartree-Fock exchange, are currently considered as the state-of-the-art.

The PBE0r exchange-correlation functional $E_\mathrm{xc}^\mathrm{PBE0r}$ \cite{Sotoudeh2017, Eckhoff2020} is a local hybrid functional derived from the frequently used PBE0 functional \cite{Perdew1996a, Adamo1999}. In contrast to PBE0, in which $\tfrac{1}{4}$ of the PBE \cite{Perdew1996} GGA exchange is replaced by Hartree-Fock exchange, PBE0r relies on on-site Hartree-Fock exchange terms $E_\mathrm{x}^\mathrm{HF,\,r}$ only, such that its form is given by
\begin{align}
E_\mathrm{xc}^\mathrm{PBE0r}=E_\mathrm{xc}^\mathrm{PBE}+\sum_{n=1}^{N}a_n\left(E_{\mathrm{x},\,n}^\mathrm{HF,\,r}-E_{\mathrm{x},\,n}^\mathrm{PBE,\,r}\right)\ .
\end{align}
Here, on-site exchange refers to those contributions, which describe the interaction between orbitals centered on the same atom. Moreover, the Hartree-Fock mixing factors $a_n$ of the $N$ atoms can be element-specific. We employ 0.07 for Li, 0.05 for O, and 0.09 for Mn \cite{Eckhoff2020}.

While the calculation of the Hartree-Fock exchange terms dominates the computational costs of hybrid functionals, in the PBE0r approach the effort is drastically reduced to a level comparable to conventional GGA functionals. However, in contrast to GGA functionals, which yield a poor description of transition metal oxides with narrow, partially filled d bands, the exact on-site exchange of PBE0r correctly splits the otherwise degenerate d band into multiplets of filled and empty orbitals leading to a more accurate description, which is also the case for Li$_x$Mn$_2$O$_4$ studied in this work.

\subsection{High-dimensional neural network potentials}

Like all machine learning potentials, HDNNPs \cite{Behler2007} provide a functional relation between the structure of the system and its potential energy. This relation crucially depends on descriptors, which fulfill the mandatory conditions of translational, rotational and permutational invariance of the potential energy surface, i.e., all equivalent representations of a system must yield the same potential energy. In HDNNPs many-body atom-centered symmetry functions \cite{Behler2011} are employed as descriptors, which ensure that these requirements are met exactly.

The local chemical environment of each atom including all neighboring atoms inside a cutoff sphere of radius $R_\mathrm{c}$ is described by a vector of atom-centered symmetry function values $\mathbf{G}$. The cutoff radius has to be sufficiently large to include all energetically relevant interactions, and typically 6 to 10\,{\AA} are used. This study employs two different types of symmetry functions, namely radial symmetry functions and angular symmetry functions, whose functional forms are given in the Supplemental Material \cite{SM}. Typically, between 20 to 200 symmetry functions are used for each atom depending on the complexity of the system.

Except for the specification of the chemical elements of the atoms no further information such as atom types, oxidation states, or predefined bonds are required making HDNNPs reactive, i.e., they are able to describe the making and breaking of bonds with the accuracy of the underlying electronic structure method. Because the dimension of the symmetry function vectors is given by the selected symmetry functions and does not dependent on the chemical environments, the symmetry function vectors are suitable as input for neural networks with fixed architecture.

Individual atomic neural networks are then constructed for each chemical element $\alpha$, which process the structural information of the geometric environment of each atom $n$ yielding its atomic energy contribution $E_n^{\alpha}$. The sum of all atomic energy contributions is the potential energy $E$ of the many-atom system containing $N_\mathrm{elements}$ elements and $N_{\rm atoms}^{\alpha}$ atoms of element $\alpha$,
\begin{align}
E=\sum_{\alpha=1}^{N_\mathrm{elements}}\sum_{n=1}^{N_\mathrm{atoms}^{\alpha}}E_n^\alpha\ . 
\end{align}
The atomic energy contributions are calculated using atomic neural networks which are standard feed-forward neural networks \cite{Haykin2008}. In the output layer a linear activation function is employed, in the neurons of the hidden layers we use the hyperbolic tangent. The functional form for the atomic energy contributions employed in our study using three hidden layers is thus given by
\begin{align}
\begin{split}
E_n^\alpha=&b_1^4+\sum_{l=1}^{n_3}a_{l1}^{34}\cdot\tanh\Bigg\{b_l^3+\sum_{k=1}^{n_2}a_{kl}^{23}\cdot\tanh\Bigg[b_k^2\\
&+\sum_{j=1}^{n_1}a_{jk}^{12}\cdot\tanh\Bigg(b_j^1+\sum_{i=1}^{n_G}a_{ij}^{01}\cdot G_{n,i}^\alpha\Bigg)\Bigg]\Bigg\}\ .
\end{split}\label{eq_energy_contribution}
\end{align}
The architecture, which consists of $n_G$ input neurons for the symmetry function values $G$, $n_1$, $n_2$, and $n_3$ neurons in the respective hidden layers, and the output layer with one neuron, is the same for all atoms of the same element. The weight parameters $a^{\rho\sigma}_{\mu\nu}$, which connect neuron $\mu$ in layer $\rho$ to neuron $\nu$ in layer $\sigma$, and bias weights $b^\sigma_\nu$, which are added to the input values of neuron $\nu$ in layer $\sigma$, are identical for all atoms of the same element. For clarity, the superscript $\alpha$ is not shown for the number of neurons per layer and for the weights in Equation \ref{eq_energy_contribution} although these quantities can be different for each element. In conclusion, for each atom in the system the individual values of the symmetry function vectors are calculated and processed in the atomic neural network of the respective element. The resulting atomic energy contributions are then added yielding the potential energy of the system.

The weight parameters of all atomic neural networks are obtained simultaneously in an iterative gradient-based optimization process, in which the errors of known potential energies and atomic force components for a set of training structures are minimized applying an adaptive, global, extended Kalman filter \cite{Kalman1960,Blank1994}. More detailed information about the HDNNP method can be found in several reviews on this topic \cite{Behler2014,Behler2015,Behler2017}.

\subsection{Initialization of the weight parameters}\label{sec_Weights}

The construction of the HDNNP requires a random initialization of the weights in the atomic neural networks as starting point for the iterative optimization. Here, we develop and implement a modification of the method proposed by Xavier \cite{Glorot2010}. Specifically, we initialize all weight parameters $a$ except for the ones connecting the last hidden layer and the output layer, $a^{34}_{\mu1}$ in Equation \ref{eq_energy_contribution}, as uniformly distributed random numbers in the range from $-$1 to 1 divided by the square root of the number of neurons in the first of the two connected layers. The bias weights $b$, except for the bias weight of the output neuron, $b_1^4$ in Equation \ref{eq_energy_contribution}, are set to 0. With this procedure proposed by Xavier \cite{Glorot2010} the input values of the hyperbolic tangent activation functions of the neurons should not be in the saturation region, which could hamper a good convergence of the fitting process. In the input layer, each symmetry function $G$ is shifted by $-G_\mathrm{mean}^\mathrm{training\ set}$ and divided by $G_\mathrm{max}^\mathrm{training\ set}-G_\mathrm{min}^\mathrm{training\ set}$ of the respective function.

For the remaining parameters, i.e., the weight parameters connecting the last hidden layer and the output layer, we developed an initialization method based on information from the reference data set. First, the range of energies $E_\mathrm{range}$ in the reference structures is calculated. Second, the mean atomic contribution to the energy $E_\mathrm{mean}$ is determined for each element by a least-squares fit. If the stoichiometry in the reference data is the same for all structures, also the mean atomic contribution averaged over all elements can be used. The weights parameters $a$ connecting the last hidden layer and the output neuron are then set to uniformly distributed random numbers in the range from $-\tfrac{E_\mathrm{range}}{2}$ to $\tfrac{E_\mathrm{range}}{2}$ divided by the square root of the number of neurons in the last hidden layer. The bias weight $b$ of the output neuron is set to $E_\mathrm{mean}$. This enables an efficient training of the HDNNP, especially for atomic neural networks with more than two hidden layers making use of a funnel-like architecture.

\section{Computational details}

\subsection{Density functional theory calculations}

The DFT reference calculations were performed using the Car-Parrinello Projector Augmented-Wave (CP-PAW) code (version from September 28, 2016) \cite{Bloechl1994, CP-PAW}. The settings of the PBE0r calculations were taken from our previous study \cite{Eckhoff2020}. Only the plane wave cutoffs for the auxiliary wave functions and for the auxiliary densities were increased to $35\,E_\mathrm{h}$ and to $140\,E_\mathrm{h}$, respectively. This yields an accuracy of 1\,meV per atom for the formation energies of Li$_x$Mn$_2$O$_4$, with $x=0,0.5,1,2$, compared to the complete basis set limit. The initial spin directions of the Mn atoms in Li$_x$Mn$_2$O$_4$ were taken from our previous study \cite{Eckhoff2020}, whereby the $\uplambda$-Mn$_2$O$_4$ spin configuration was used for $0\leq x\leq\tfrac{1}{8}$ and the LiMn$_2$O$_4$ spin configuration for $\tfrac{1}{4}\leq x\leq1$ corresponding to the energetic preference.

The atomic spins were calculated by a mapping the spin density onto the one-center expansions of the partial waves. For each atom the contributions inside a cutoff sphere were added. The cutoff radii were set to 1.2 times the covalent radii of the corresponding atoms which are given in our previous study \cite{Eckhoff2020}. 

The D3 dispersion corrections using Becke-Johnson damping were calculated by the DFT-D3 software (version from June 14, 2016) \cite{Grimme2010, Grimme2011}. The settings given for the HSE06 functional were applied as described in our previous benchmark \cite{Eckhoff2020}.

The nudged elastic band calculations \cite{Mills1994, Jonsson1998} were performed using CP-PAW and the Python module of the Atomic Simulation Environment (ASE) \cite{Larsen2017} employing a self-written interface.

\subsection{Construction of the reference data set}

While the functional form of the HDNNP is very flexible, which allows to represent the reference DFT data very accurately, it does not have a physical foundation. Thus, the correct topology of the potential energy surface has to be ``learned'' exclusively from the information provided in the reference data set. Therefore, the quality of the reference data set, which has to cover the configuration space relevant for the intended simulations, is essential for the reliability of the HDNNP.

Three important points have to be considered for the reference data set. First, the reference method has in itself to be accurate enough to describe the material of interest. In the present case, we confirmed that the PBE0r-D3 DFT functional provides a reliable description of Li$_x$Mn$_2$O$_4$ as reported in our previous benchmark \cite{Eckhoff2020}.

Second, the amount and diversity of information in the reference data set has to be sufficient to represent the relevant part of the configuration space reliably. In the past, we have developed a detailed protocol based on active learning to systematically identify and add important configurations to the reference data set in a self-consistent process \cite{Artrith2012}. This process is accompanied by a careful multi-step validation process to ensure that all configurations visited in our simulations are well-represented in the database and thus correctly described by the HDNNP. The iterative learning process is summarized below as starting point in the discussion of specific challenges in the construction of the reference data set for a system with a complex electronic structure as Li$_x$Mn$_2$O$_4$. 

Third, the reference data set has to consistently represent one potential energy surface determined employing one reference method. One electronic state -- typically the ground state -- of the system can be considered as otherwise the potential energy would not be a well-defined function of the atomic structure. Therefore, we only consider ground state spin configurations and one consistent arrangement of spin directions. As many arrangements in Li$_x$Mn$_2$O$_4$ are almost degenerate within the accuracy of the DFT reference calculations \cite{Eckhoff2020}, the arrangement of spin directions has only a marginally small influence on the energy.

For an initial set of configurations, experimentally determined structures can be used. In order to include information about thermal distortions, the atomic positions and lattice vectors can be modified randomly within a reasonable range. In this study we displaced the atomic positions inside spheres with radii of up to $0.2\,\mathrm{\AA}$ and changed the lattice vectors within a range of at most $\pm 3\%$. The energy and force components of each system have then been determined in electronic structure calculations. With this initial data, a preliminary HDNNP can be constructed that allows to perform MD simulations to further explore the configuration space guided by the atomic forces. By using two different HDNNPs or even a larger ensemble, obtained for example using different initial weights and neural network architectures but trained on the same reference data set, missing configurations can be identified based on the deviations in their predicted energies and forces. Deviations above a predefined threshold indicate structures being too distant from the available training data, which should then be recomputed by DFT and added to the reference set to improve the HDNNP. The HDNNP autonomously explores and learns the configuration space of the investigated system by iteratively repeating this process.

However, in particular in the early stages of this process, unphysical trajectories can emerge if configurations are visited in the simulations, which are too different from the training set of the HDNNP. Removing structures with unphysically short nearest-neighbour distances or too high energies and forces, which do not occur at the temperatures and pressures of interest, is advisable to restrict the reference data set to include only reasonable structures. Moreover for unphysically close atoms, electronic structure calculations can fail to converge to the correct ground-state energy. Even if electronic structure calculations are carefully checked for a good convergence behavior, they still can get stuck in metastable electronic states, which is difficult to detect. In particular for Li$_x$Mn$_2$O$_4$ we observed this phenomenon due to its complex electronic structure and variety of electronic states that are energetically close to the ground state. Therefore, we followed two strategies to clean the reference data set from incorrect electronic structure data:

1.\ In Li$_x$Mn$_2$O$_4$ the number of Mn$^\mathrm{III}$ ions is equal to the number of Li ions. If the result of an electronic structure calculation differs from this relation, it does not provide the correct electronic ground state. In order to check this relation we calculated the absolute values of the atomic spins using DFT. hs-Mn$^\mathrm{III}$ ions have an absolute atomic spin of $2\,\hbar$, while the spin of Mn$^\mathrm{IV}$ ions is $\tfrac{3}{2}\,\hbar$. This allows us to count the number of Mn$^\mathrm{III}$ and Mn$^\mathrm{IV}$ ions to identify calculations that did not yield the right electronic ground state. 
However, due to thermal fluctuations, intermediate structures exists where the Jahn-Teller distortion of an octahedron of an initial hs-Mn$^\mathrm{III}$ ion decreases while that of an adjacent close-to-ideal octahedron increases. As a consequence, a transition exists in which the e$_\mathrm{g}$ electron of a Mn$^\mathrm{III}$O$_6$ octahedron can hop to an adjacent Mn$^\mathrm{IV}$O$_6$ octahedron. In the transition state the e$_\mathrm{g}$ electron is shared between both sites as in Zener polarons. 
This information is important for the potential energy surface and must not be removed from the reference data set. However, structures including Mn ions with atomic spins outside the interval between $\tfrac{3}{2}$ and $2\,\hbar$, structures with too few or too many Mn$^\mathrm{III}$ ions considering the transition states, or structures including Mn$^\mathrm{III}$ ions in octahedra corresponding to a Mn$^\mathrm{IV}$O$_6$ geometry and vice versa, are not suitable for the reference data set because they do not describe the electronic ground state.

2.\ Instead of using physical properties to identify incorrect data points, the second method is based on the predictive power of HDNNPs. If the electronic structure of a system is not properly converged, the energy will be different from the correct ground state energy. Unfortunately, in most cases these energies cannot be identified as outliers in the data set because thermal distortions or different stoichiometries lead to larger energy variations. Still, the training process of the HDNNP reveals these suspicious energies because they do not fit to the general trend. The goal of the fitting process of the weight parameters is to minimize the root mean squared error of the energies. Since most of the structures of the training set are correct, these data points dominate the fitting process, and the minority of problematic structures, which has energies and forces being incompatible with the majority of the data, is less well described. In order to distinguish between less accurately fitted but correct data points and structures with a problematic electronic structure, we can plot the differences between predicted energies and reference energies of one HDNNP against the differences of another HDNNP, both fitted on the same complete reference data set, but using different initial weights and neural network architectures. If data points show the same large error, irrespective of the specific HDNNP, it is very likely that the electronic structure did not converge correctly and the data point should be removed. On the other hand, energies which are not well represented in just one of the two HDNNPs are just poorly fitted. Such deviations reveal that the general quality of the HDNNP has to be improved, but the corresponding data should be kept in the data set.

Further details about the construction of the reference data set can be found in various reviews \cite{Behler2014,Behler2015,Behler2017}.

\subsection{Construction of the high-dimensional neural network potential}

The construction of the HDNNP was carried out using the RuNNer code (version from August 22, 2019) \cite{Behler2015, Behler2017, RuNNer}. The atomic neural networks consist of an input layer with $n_G^\alpha$ neurons depending on the specific element, three hidden layers with 20, 15, and 10 neurons, respectively, and an output layer with one neuron. The employed symmetry functions with a cutoff radius of $R_\mathrm{c}=12\,a_0$, with $a_0$ being the Bohr radius, are given in the Supplemental Material \cite{SM}. The HDNNP was trained on the formation energies, i.e., on total DFT energies minus the sums of the energies of the atoms in their reference structures, which are body centered cubic Li, gaseous O$_2$, and $\upalpha$-Mn. Further, the DFT atomic force components ($F_x$, $F_y$, $F_z$) of the reference structures have been used for training the HDNNP. 90{\%} of the reference data were used to determine the weight parameters. With the remaining 10{\%} the reliability of the HDNNP for unknown structures was tested. The initialization of the weights in the atomic neural networks is discussed in Section \ref{sec_Weights}. Further details about the training procedure are given in the Supplemental Material \cite{SM}.

\subsection{Simulations}

The MD simulations were performed using the Large-scale Atomic/Molecular Massively Parallel Simulator (LAMMPS) (version from August 7, 2019) \cite{Plimpton1995, LAMMPS}. The neural network potential package (n2p2) (version from December 9, 2019) \cite{n2p2} was included in order to use HDNNPs with LAMMPS. The Nos\'{e}-Hoover \cite{Nose1984, Hoover1985} thermostat and barostat were applied to run simulations in the isothermal-isobaric ($NpT$) ensemble. The coupling constants were set to 0.05\,ps and 0.5\,ps, respectively. The time step of the MD simulations was 0.5\,fs except for the microcanonical ($NVE$) simulations to calculate the phonon density of states with the aid of the Python package pwtools \cite{pwtools}, for which it was reduced to 0.25\,fs. All $NpT$ simulations were performed at a pressure of $p=1\,\mathrm{bar}$. The simulation cell vector angles were fixed at 90\,$^\circ$. Thermodynamic data were calculated after an initial equilibration period of 1\,ns.

The basin-hopping Monte Carlo \cite{Wales1997} calculations also used LAMMPS and n2p2 for the geometry optimizations employing the conjugate gradient algorithm. Only Monte Carlo steps of the Li ions between the tetrahedral sites occupied in LiMn$_2$O$_4$ were allowed. The positions of all other atoms were only changed during the geometry optimizations. This grid-based basin-hopping Monte Carlo (GBHMC) approach allows an efficient sampling of different Li distributions inside Li$_x$Mn$_2$O$_4$.

\section{Experimental material and methods}

To experimentally resolve the effects of changing the Li content and temperature on the LiMn$_2$O$_4$ crystal structure and lattice constants, in situ XRD techniques have been applied to nanoparticle powder specimens (Sigma-Aldrich, single synthesis charge, purity $>99\%$), where the measured XRD signals represent the volume average over many particles. The mean particle diameter was identified to be $(44\pm14)$\,nm using transmission electron microscopy and scanning electron microscopy \cite{Baumung2019a}. The particles mainly consist of truncated octahedra and some truncated rhombic dodecahedra \cite{Baumung2019a}. An XRD diffractogram (see Supplemental Material \cite{SM, Izumi2007, Momma2011}) of the as purchased material fits to the expected cubic spinel structure with a lattice constant of 8.234(2)\,{\AA}. Two additional cubic phases with a volume share of less than 1{\%} have also been detected. Applying the Scherrer equation to the LiMn$_2$O$_4$ peak widths, a volume averaged crystallite size of $(61\pm9)$\,nm has been found, which is larger than the unweighted value obtained from imaging, as expected.

In situ XRD diffractograms were recorded during (de)lithiation in the range of $0.48\leq x\leq2$ with a specifically designed battery cell at CIC energiGune in Vitoria-Gasteiz, Spain. The cell consists of a Be window, which is coated with a thin Al foil on the inner surface, followed by the positive electrode material consisting of 85\,wt{\%} LiMn$_2$O$_4$ and 15\,wt{\%} carbon black, and then by a separator and a metallic Li negative electrode. 1\,mol\,l$^{-1}$ LiPF$_6$ in EC/DMC was used as electrolyte. The cell was assembled in a glove box and then cycled from 3.0\,V to 4.5\,V and back again before it was held for 20\,h at open circuit voltage. This was followed by (de)lithiation of Li$_x$Mn$_2$O$_4$ from $x=1$ to $x=0.48$ and back to $x=1$ with a set rate of C/30. A series of XRD measurement scans were performed during voltage cycling, each with step sizes of $\Delta2\theta=0.02\,^\circ$ in the range $40\,^\circ\leq2\theta\leq55\,^\circ$, and were synchronized with the electrochemical measurements of transferred charge and cell potential. The values of $2\theta$ were corrected for the offset of the electrode away from the goniometer axis. The positions of the most intense peaks (311) and (400) were fitted with pseudo Voigt functions, from which the lattice constant $a$ has been calculated using Bragg's law and $d_{hkl}=\tfrac{a}{\sqrt{h^2+k^2+l^2}}$. The displayed data were collected during the discharge (Li intercalation) to $x=1$ because of overpotential and current loss during charging. 

Temperature dependent X-ray diffractograms were taken between $T=100$\,K and 305\,K with a step size of $\Delta T=30$\,K using a Bruker Smart APEX II Goniometer with an area detector covering an angular range of $0\,^\circ<2\theta<35\,^\circ$, $\Delta 2\theta=0.095\,^\circ$. Close to the phase transition temperature, from 260\,K to 295\,K, a step size of $\Delta T=5$\,K was used. The LiMn$_2$O$_4$ powder specimen was compacted into a glass capillary, which was placed in a nitrogen gas flow cryostat with a temperature sensor. The specimen was held for 30\,min after reaching the initial nitrogen gas setpoint of 100\,K to allow for thermal equilibration prior to taking the first diffractogram. Thermal equilibration times of 10\,min were allowed for each subsequent temperature step. The recorded polycrystalline diffracted ring pattern was then azimuthally integrated. Although the small peak splitting due to the cubic to orthorhombic transition has been observed using high resolution synchrotron and neutron diffraction experiments in powder samples \cite{Rodriguez-Carvajal1998, Massarotti1999, Piszora2004} and using single crystal XRD \cite{Akimoto2000, Akimoto2004}, it could not be resolved in our XRD experiments of polycrystalline powder sample, so that the observed peaks were fit with a cubic structure using the procedure described above.

\section{Results and discussion}

\subsection{High-dimensional neural network potential}

The reference data set of the HDNNP contains 15228 Li$_x$Mn$_2$O$_4$ bulk structures, with $0\leq x\leq2$. These structures sample the configuration space of Li$_x$Mn$_2$O$_4$ up to a temperature of about 500\,K. 13669 reference structures have been used for training, the remaining 1559 structures form the test set. In total, 682431 atomic environments and 2047293 force components are thus available for training and 77911 atomic environments and 233733 force components for testing. More detailed information about the number of structures for each composition are given in the Supplemental Material \cite{SM}.

The formation energies cover a range from $-$2.15 to $-$1.61\,eV\,atom$^{-1}$. The root mean squared error (RMSE) of the energies is 1.8\,meV\,atom$^{-1}$ for the training set and 2.2\,meV\,atom$^{-1}$ for the test set. The maximal error of the energies is 9.9\,meV\,atom$^{-1}$ for the training set and 10.9\,meV\,atom$^{-1}$ for the test set. 98.00{\%} of the structures in the training set and 95.25{\%} of the structures in the test set have an error smaller than 5\,meV\,atom$^{-1}$.

\begin{figure*}[tb!]
\centering
\includegraphics[width=\textwidth]{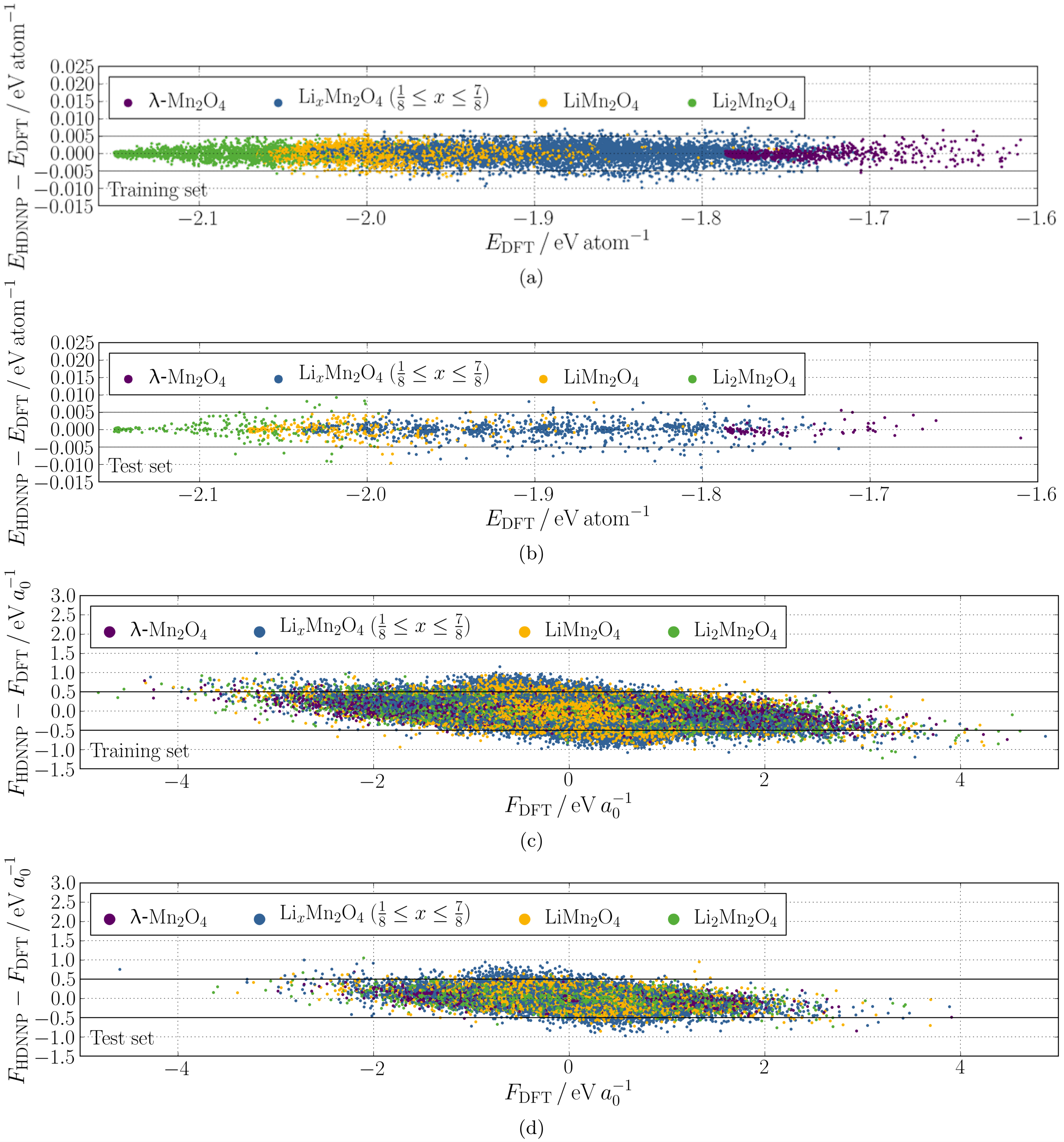}
\caption{Signed errors of the formation energies for the (a) training set and (b) test set and of the force components for the (c) training set and (d) test set of the HDNNP as a function of the respective DFT values. The data points are colored according to the Li content of the underlying structures. Inside the black lines the error is smaller than 5\,meV\,atom$^{-1}$ for the energies and 0.5\,eV\,$a_0^{-1}$ for the forces, respectively.}
\label{fig_NNP_quality}
\end{figure*}

The largest absolute force component in the reference data set is 4.87\,eV\,$a_0^{-1}$. The RMSE of the force components is 0.108\,eV\,$a_0^{-1}$ for the training set, while it is 0.107\,eV\,$a_0^{-1}$ for the test set. The maximum errors are 1.50 and 1.05\,eV\,$a_0^{-1}$ for the training and test set, respectively. Only 0.29{\%} of the 2047293 force components in the training set show an absolute error greater than 0.5\,eV\,$a_0^{-1}$. The respective fraction for the test set is 0.29{\%} of the 233733 force components. Figures \ref{fig_NNP_quality} (a) to (d) show the deviations between HDNNP and DFT formation energies and force components for both data sets as a function of the respective DFT values. 

The reliability of the HDNNP was tested in 10\,ns $NVE$ MD simulations starting from Li$_x$Mn$_2$O$_4$ structures with different Li content in the range $0\leq x\leq2$ equilibrated at 500\,K in $NpT$ MD simulations. The energy conservation in the $NVE$ simulations is very good. Drifts of the total energy are only in the order of $10^{-6}$\,eV\,atom$^{-1}$\,ns$^{-1}$. The configuration space covered by the training data was not exceeded during any simulation, i.e., no extrapolation was detected.

\subsection{Validation on density functional theory and experimental data}\label{sec_Validation}

In order to assess the quality of the HDNNP for Li$_x$Mn$_2$O$_4$ we will first compare several physical properties determined using the HDNNP with results that are accessible by PBE0r-D3 DFT calculations. The PBE0r-D3 DFT functional was also used for the energy and force calculations of the reference data set. Therefore, the HDNNP results should match the PBE0r-D3 DFT data.

Li$_x$Mn$_2$O$_4$ is used as positive electrode material in batteries, and Li intercalation and deintercalation are crucial processes. Consequently, the Li diffusion barrier is an important property of this material. The barrier can be calculated using the nudged elastic band method. PBE0r-D3 DFT provides a barrier of 0.52\,eV in the Li$_{0.875}$Mn$_2$O$_4$ unit cell, which is in very good agreement with previous experimental NMR data yielding $(0.5\pm0.1)\,\mathrm{eV}$ \cite{Verhoeven2001}. The HDNNP predicts a barrier of 0.55\,eV for the same minimum energy path matching closely the DFT reference data as well as experiment.

Apart from Li diffusion, the electrochemical potential is very important for battery applications and degradation of Li$_x$Mn$_2$O$_4$ during electrocatalysis of oxygen \cite{Baumung2019}. In the case of Li$_x$Mn$_2$O$_4$ it depends on the Li content. From experiments it is known that the standard electrochemical potential of a Li/Li$_x$Mn$_2$O$_4$ cell exhibits a plateau at about 4.1\,V for $0<x<\tfrac{1}{2}$, 4.0\,V for $\tfrac{1}{2}<x<1$ \cite{Xia1996, Baumung2019}, and 3.0\,V for $1<x<2$ \cite{Peramunage1998} leading to differences between these plateaus of 0.1\,eV and 1.0\,eV, respectively. Following the approach of our previous study \cite{Eckhoff2020}, we obtain 0.05 and 0.95\,eV using the PBE0r-D3 formation energies. The HDNNP yields similar values of 0.02 and 1.07\,eV for 0\,K. HDNNP MD simulations under standard ambient conditions (298\,K, 1\,bar) are also feasible. The electrochemical potential differences between the plateaus calculated from the mean potential energy of 40\,ns $NpT$ MD simulations of different Li loads are 0.02\,eV and 1.05\,eV. These results show that the thermal energy is of minor importance for the electrochemical potential of Li$_x$Mn$_2$O$_4$, in agreement with a previous study \cite{Liu2016a}. The results are also in good agreement with experimental data \cite{Ohzuku1989}.

For dynamical studies it is important that the HDNNP correctly represents the vibrational motions of the system. To validate the vibrational fingerprint of the simulations the calculated phonon density of states \cite{Lee1993} is compared with experimental Fourier-transform infrared and Raman spectra \cite{Rougier1998, Ramana2005, Li2009, Helan2010} in the Supplemental Material \cite{SM}.

\subsection{Structural changes due to lithium intercalation}

Li intercalation and deintercalation during discharge and charge of a battery includes structural expansion and contraction of Li$_x$Mn$_2$O$_4$. This behaviour can be studied in theory as well as in experiment. XRD experiments yield reliable and precise values for the lattice constants. However, the data are averaged over time and space hiding information about individual atomic environments. On the other hand, simulations rely on appropriate structural models, but they reveal the underlying processes on an atomic scale with femtosecond resolution. Therefore, we will combine the advantages of both approaches by validating the HDNNP predictions first on experimental and DFT data and then obtaining a detailed atomic scale understanding from our simulations.

We start the investigation at 0\,K to confirm the HDNNP results for the optimized structures using PBE0r-D3 DFT reference data. The conventional Li$_x$Mn$_2$O$_4$ unit cells are still accessible by DFT. In order to find the global minimum Li distributions, with $\tfrac{1}{4}\leq x\leq\tfrac{3}{4}$, we performed GBHMC simulations using the HDNNP. The low-energy structures were then reoptimized by DFT. The PBE0r-D3 results indicate that there is an effective repulsion between the Li ions on neighboring tetrahedral sites, leading to only every other tetrahedral site being populated for $0<x\leq\tfrac{1}{2}$ in the most stable structures. Neighboring tetrahedral sites only become occupied by Li for $x>\tfrac{1}{2}$. This change in occupation is presumably the reason for the plateaus of the electrochemical potentials in the region $0<x<\tfrac{1}{2}$ and $\tfrac{1}{2}<x<1$ as reported previously \cite{vanderVen2000, Baumung2019a}.

\begin{figure}[tb!]
\centering
\includegraphics[width=\columnwidth]{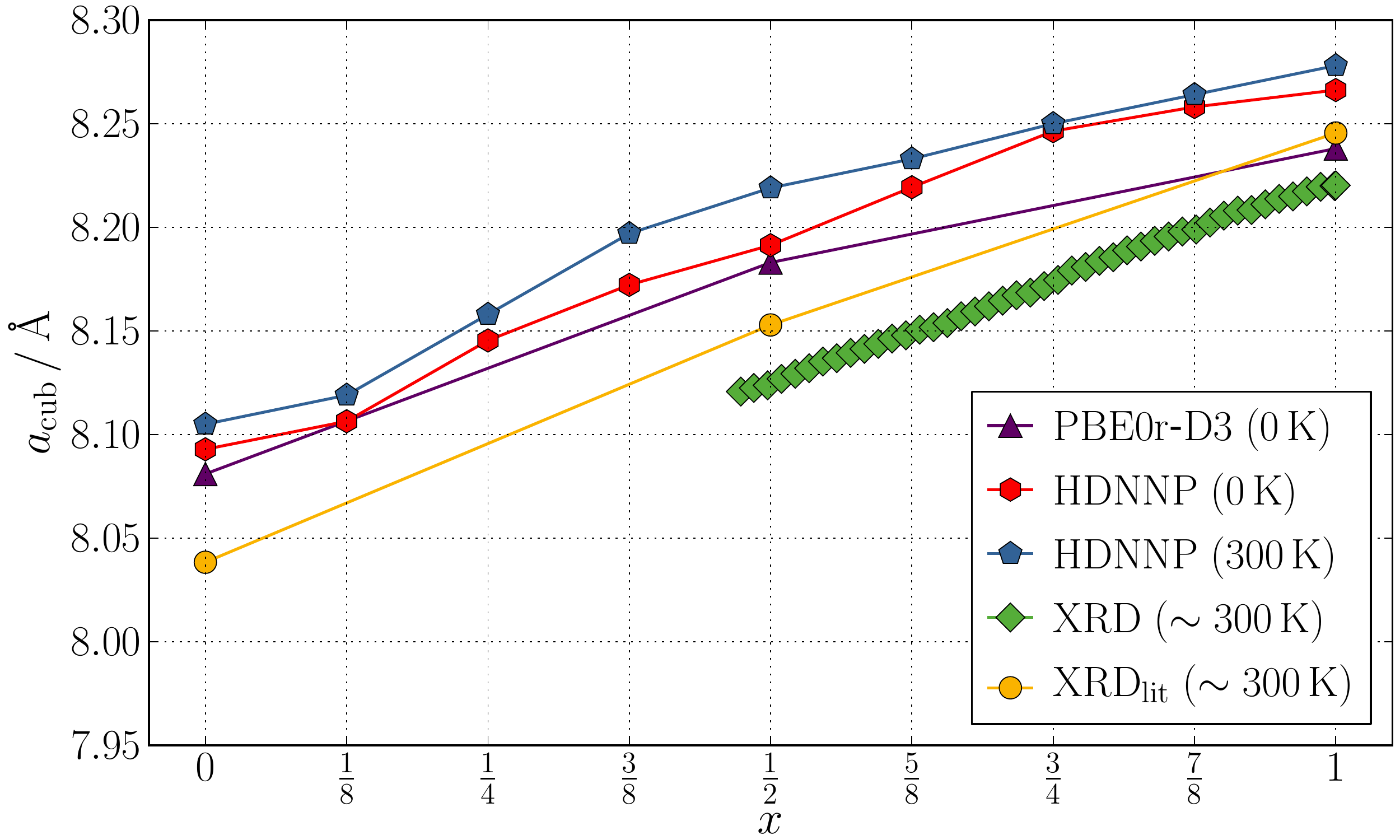}
\caption{Cubic lattice constant $a_\mathrm{cub}$ (Equation \ref{eq_cubic}) as a function of the Li content $x$ determined by PBE0r-D3 and the HDNNP at 0\,K as well as by the HDNNP and XRD at about 300\,K. The XRD$_\mathrm{lit}$ data have been taken from References \cite{Takahashi2003, Bianchini2015, Akimoto2004}.}
\label{fig_x-a_111_300}
\end{figure}

The PBE0r-D3 equilibrium volumes of the Li$_x$Mn$_2$O$_4$ unit cells, with $x=0$, 0.5, 1, and 2, were determined using the Birch-Murnaghan equation of state \cite{Murnaghan1937, Birch1947}. The HDNNP was used to optimize the atomic positions as well as the lattice vectors of the Li$_x$Mn$_2$O$_4$ unit cells, with $x=0$ to 1 in steps of $\tfrac{1}{8}$ and $x=2$. The lattice angles of 90\,$^\circ$ were kept constant in these optimizations. In order to compare the results with experiment, Figure \ref{fig_x-a_111_300} shows the averaged cubic lattice constant (geometric mean),
\begin{align}
a_\mathrm{cub}=(a_x\cdot a_y\cdot a_z)^{\tfrac{1}{3}}\ . \label{eq_cubic}
\end{align}
The minimum structures predicted by the HDNNP are given in the Supplemental Material \cite{SM} along with the bulk moduli, which have also been determined and compared to DFT and experiment \cite{Lin2011}.

The HDNNP lattice constants match the PBE0r-D3 DFT data very well. The cubic lattice constants agree within 0.5{\%} for $x=0$, 0.5, and 1. The PBE0r-D3 results show that the lattice expands by 0.102\,$\mathrm{\AA}$ in the range from $x=0$ to 0.5 and by 0.055\,$\mathrm{\AA}$ from $x=0.5$ to 1. The HDNNP predicts similar expansions of 0.105\,$\mathrm{\AA}$ and 0.075\,$\mathrm{\AA}$, respectively. The same trend is present in published XRD data of $\uplambda$-Mn$_2$O$_4$ (at room temperature) \cite{Takahashi2003}, Li$_{0.5}$Mn$_2$O$_4$ (at 293\,K) \cite{Bianchini2015}, and LiMn$_2$O$_4$ (at 330\,K) \cite{Akimoto2004} with 0.114\,$\mathrm{\AA}$ and 0.093\,$\mathrm{\AA}$, respectively, although the expansion is less dependent on the absolute Li content. The optimized HDNNP results at 0\,K reveal that the increase of the cubic lattice constant is slightly smaller if a Li fcc sublattice is already almost filled. As a consequence, a small dip at $x=0.5$ is visible and the slope is reduced closer to $x=1$ in Figure \ref{fig_x-a_111_300}. Our in-situ XRD data yield a linear increase of 0.098\,$\mathrm{\AA}$ in the range from $x=0.5$ to 1. This linear trend and the extrapolated result of the lattice constant at $x=0$ of 8.027\,$\mathrm{\AA}$ are in good agreement with previous data \cite{Hunter1981, Ohzuku1989, Takahashi2003}.

The optimized lattice constants of Li$_2$Mn$_2$O$_4$ are predicted to be 5.641, 5.641, and 9.227\,$\mathrm{\AA}$ using PBE0r-D3 and 5.614, 5.614, and 9.322\,$\mathrm{\AA}$ using the HDNNP. Therefore, the agreement is within 1.0{\%} although the ratio of the lattice constants in the PBE0r-D3 calculations has been restricted to the experimental value for the $a_z$ over $a_x$ ratio, while it was not restricted in the HDNNP optimizations. The lattice constants are also in good agreement with XRD measurements yielding 5.650, 5.650, and 9.242\,$\mathrm{\AA}$ at room temperature \cite{Mosbah1983}.

For a realistic comparison with experiment, simulations at finite temperatures must be carried out. Therefore, 40\,ns $NpT$ MD simulations of Li$_x$Mn$_2$O$_4$ unit cells at 300\,K were performed using the HDNNP. The resulting $a_\mathrm{cub}(x)$ graph in the range from $0\leq x\leq1$ is shifted to larger lattice constant values and is generally smoother than the HDNNP results at 0\,K (Figure \ref{fig_x-a_111_300}). The increase of the lattice constant as a function of Li content is similar to that of the XRD data with a change of 0.122\,$\mathrm{\AA}$ in the range $0\leq x\leq0.5$ and a change of 0.060\,$\mathrm{\AA}$ in the range $0.5\leq x\leq1$. The finite temperature simulations are systematically offset relative to the XRD data by around 0.1\,$\mathrm{\AA}$. On the one hand, this deviation of about 1{\%} is caused by an overestimation of the lattice constants by the underlying PBE0r-D3 data compared to reality. On the other hand, it can either be explained by an error in experimental $x$ determination or by stoichiometric deviations of the sample material from the ideal Li$_x$Mn$_2$O$_4$ formula. In the case of our XRD data, the value for $x$ is determined from the integrated in situ cell current and is only accurate to within $\pm0.02$. Small changes of the parameters for the sample preparation process can easily influence the stoichiometry and thereby the lattice constant \cite{Xia2001}.

\begin{figure}[tb!]
\centering
\includegraphics[width=\columnwidth]{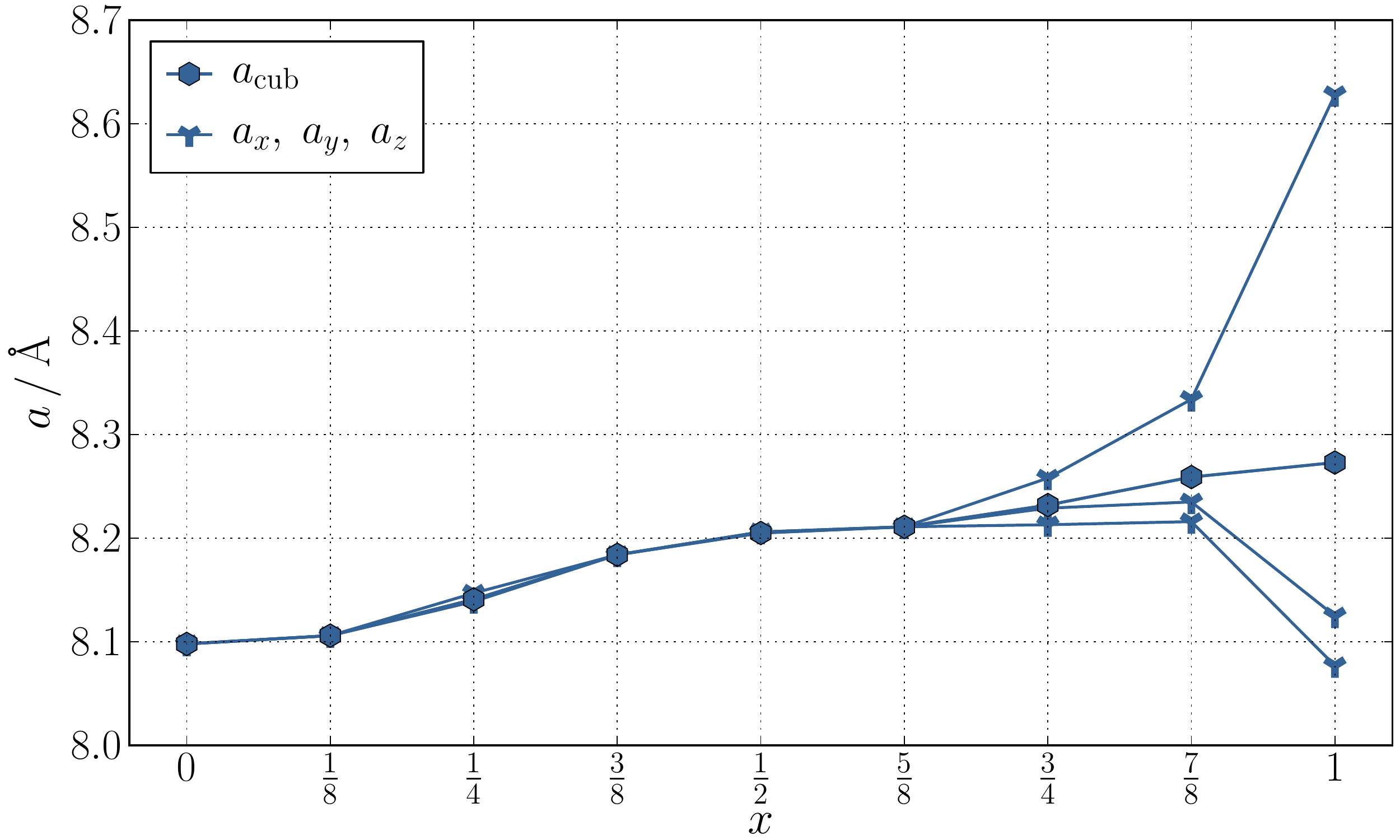}
\caption{Lattice constants $a_\mathrm{cub}$ and $a_{x,y,z}$ as a function of the Li content $x$ determined by the HDNNP in a 40\,ns $NpT$ simulation at 150\,K.}
\label{fig_x-a_111_150}
\end{figure}

Using the HDNNP we can investigate the system also at lower temperatures. Figure \ref{fig_x-a_111_150} shows the cubic lattice constant $a_\mathrm{cub}$ as well as the individual lattice constants $a_x$, $a_y$, and $a_z$, each averaged over a 40\,ns $NpT$ MD simulation at 150\,K. At Li contents $x\leq\tfrac{5}{8}$ the time averaged values of $a_x$, $a_y$, and $a_z$ are virtually equal to the cubic lattice constant $a_\mathrm{cub}$. However, at larger Li contents the time averaged structure remains orthorhombic. This is in agreement with experimental measurements at low temperatures, which yield an orthorhombic structure with lattice constants 24.750, 24.801, and 8.190\,{\AA} at 130\,K for $x=1$ \cite{Akimoto2004} corresponding to a distorted $3\times3\times1$ supercell of the unit cell at higher temperatures. The less pronounced orthorhombicity in experiment will be discussed in Section \ref{sec_phase_transition}. For $x=1$ the lattice constants at 150\,K are almost equal to the optimized results at 0\,K, with only a slight increase due to thermal expansion.

\begin{figure}[tb!]
\centering
\includegraphics[width=\columnwidth]{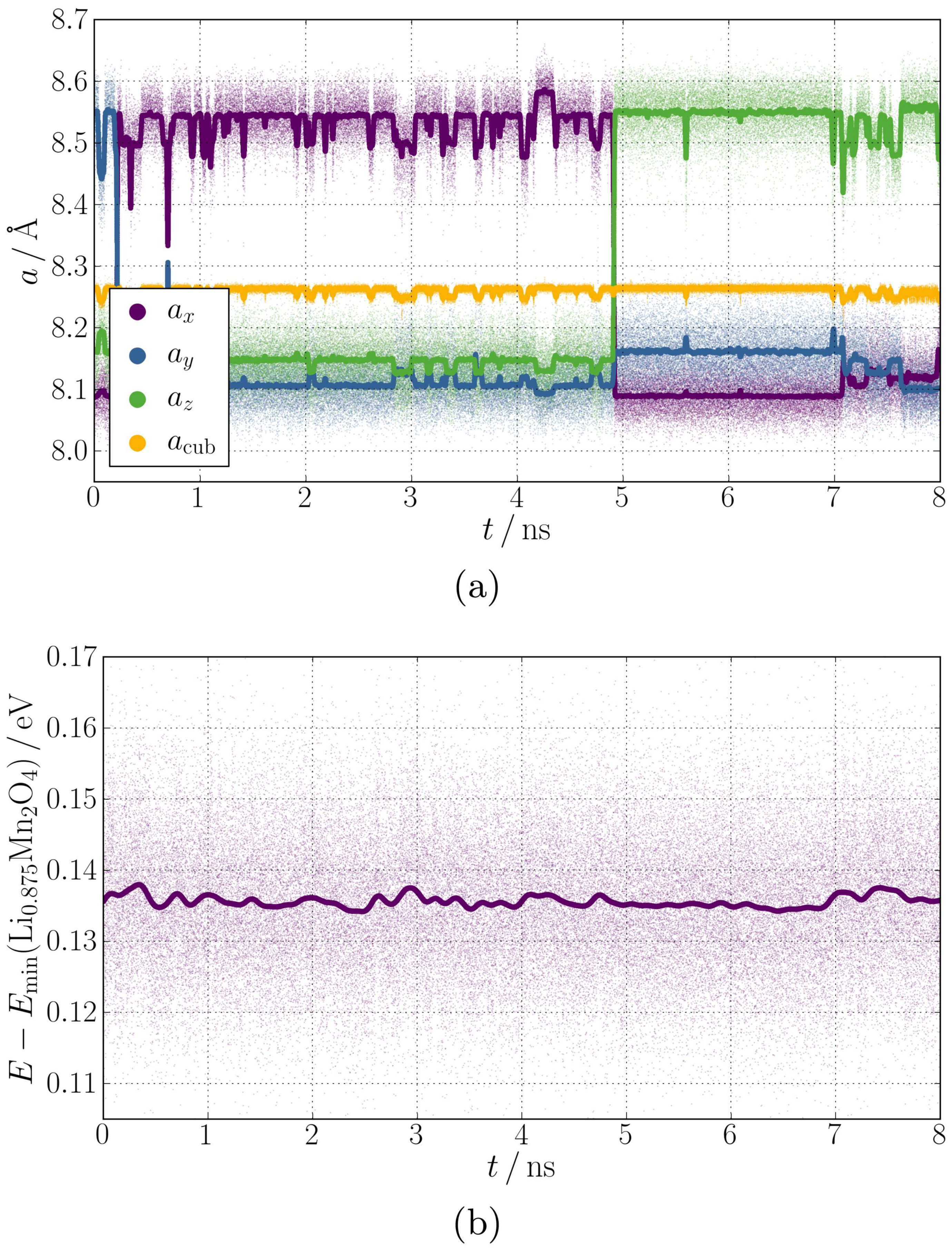}
\caption{(a) Lattice constants $a_{x,y,z}$ as well as $a_\mathrm{cub}$ and (b) potential energy $E-E_\mathrm{min}(\mathrm{Li}_{0.875}\mathrm{Mn}_2\mathrm{O}_4)$ per formula unit as a function of the time $t$ determined by the HDNNP at 150\,K for Li$_{0.875}$Mn$_2$O$_4$. $E_\mathrm{min}(\mathrm{Li}_{0.875}\mathrm{Mn}_2\mathrm{O}_4)$ is the potential energy of the HDNNP optimized Li$_{0.875}$Mn$_2$O$_4$ structure. The data were collected in 1\,fs intervals in the simulation, averaged over 100\,fs, and finally smoothed by the Kolmogorov-Zurbenko filter \cite{Zurbenko1986}. The averaged data of each 100\,fs are shown as scatter plot. The displayed data are a representative interval of a 40\,ns MD simulation of the unit cell.}
\label{fig_t-aE_7_111_150}
\end{figure}

In order to investigate why the time average leads to a cubic structure at lower Li contents, we plotted the lattice constants $a_x$, $a_y$, $a_z$ and $a_\mathrm{cub}$ as well as the potential energy $E$ as a function of the time $t$ for the Li$_{0.875}$Mn$_2$O$_4$ unit cell at 150\,K in Figures \ref{fig_t-aE_7_111_150} (a) and (b). We chose Li$_{0.875}$Mn$_2$O$_4$ because for this Li content the time averaged lattice constants at 150\,K are between the cubic and the optimized orthorhombic crystal structure (see Supplemental Material \cite{SM}). The data were collected each femtosecond of the $NpT$ MD simulation and then averaged over 100\,fs intervals. These averaged data were finally smoothed by the Kolmogorov-Zurbenko filter \cite{Zurbenko1986}, which uses $k$ iterations of a moving average filter with window width $m$. We applied this low-pass filter with the parameters $k=3$ and $m=101$ for the lattice constants and $m=1001$ for the potential energy.

Figure \ref{fig_t-aE_7_111_150} (a) reveals that the distortion of the lattice constants is still similar to the optimized structure. However, the elongation of the cell changes its spatial direction, leading to a cubic lattice as an average over time. This is caused by the Jahn-Teller distortions of the Mn$^\mathrm{III}$O$_6$ octahedra, which are dynamically fluctuating, both due to spatial changes in the orientation of the distorted octahedra and due to electron transfers converting Mn$^\mathrm{III}$ to Mn$^\mathrm{IV}$ and vice versa, as will be discussed below in more detail. At lower $x$ the fluctuations become even faster but the structures are still orthorhombic at any point in time like the minimized structures for $x>0$. Figure \ref{fig_t-aE_7_111_150} (b) shows that the overall changes of the potential energy are quite small, which explains why the fluctuations already happen at low temperatures.

For lower Li contents, fewer Jahn-Teller distorted Mn$^\mathrm{III}$O$_6$ octahedra are present, and the orthorhombic structure becomes less favorable. As a consequence, the transition temperature of the ordered orthorhombic phase to the disordered cubic phase decreases. Consistently, in experiment it has been found that a higher O content, i.e. fewer Mn$^\mathrm{III}$O$_6$ octahedra, lead to a decrease of the transition temperature to the cubic phase \cite{Tachibana2003}. This is because a higher oxygen-to-manganese ratio results in higher oxidation states of manganese, i.e. more Mn$^\mathrm{IV}$. Consequently, oxygen vacancies have the opposite effect, yielding more Mn$^\mathrm{III}$, just as for an increased Li content. In fact, in Li rich spinels, Li$_{1+x}$Mn$_{2-x}$O$_4$, in which Mn is partially replaced by Li on octahedral sites, the cubic to orthorhombic transition can be suppressed to at least below $T=20\,$K \cite{Tabuchi1997}.

\subsection{Role of temperature}

In order to investigate the transition from an orthorhombic to a cubic structure, we performed 40\,ns $NpT$ MD simulations of the LiMn$_2$O$_4$ unit cell in the range from 10 to 500\,K. These were compared with experimental measurements on LiMn$_2$O$_4$ which yield an orthorhombic crystal structure with lattice constants 24.750, 24.801, and 8.190\,{\AA} at 130\,K and a cubic crystal structure with a lattice constant of 8.246\,{\AA} at 330\,K \cite{Akimoto2004}.

\begin{figure}[tb!]
\centering
\includegraphics[width=\columnwidth]{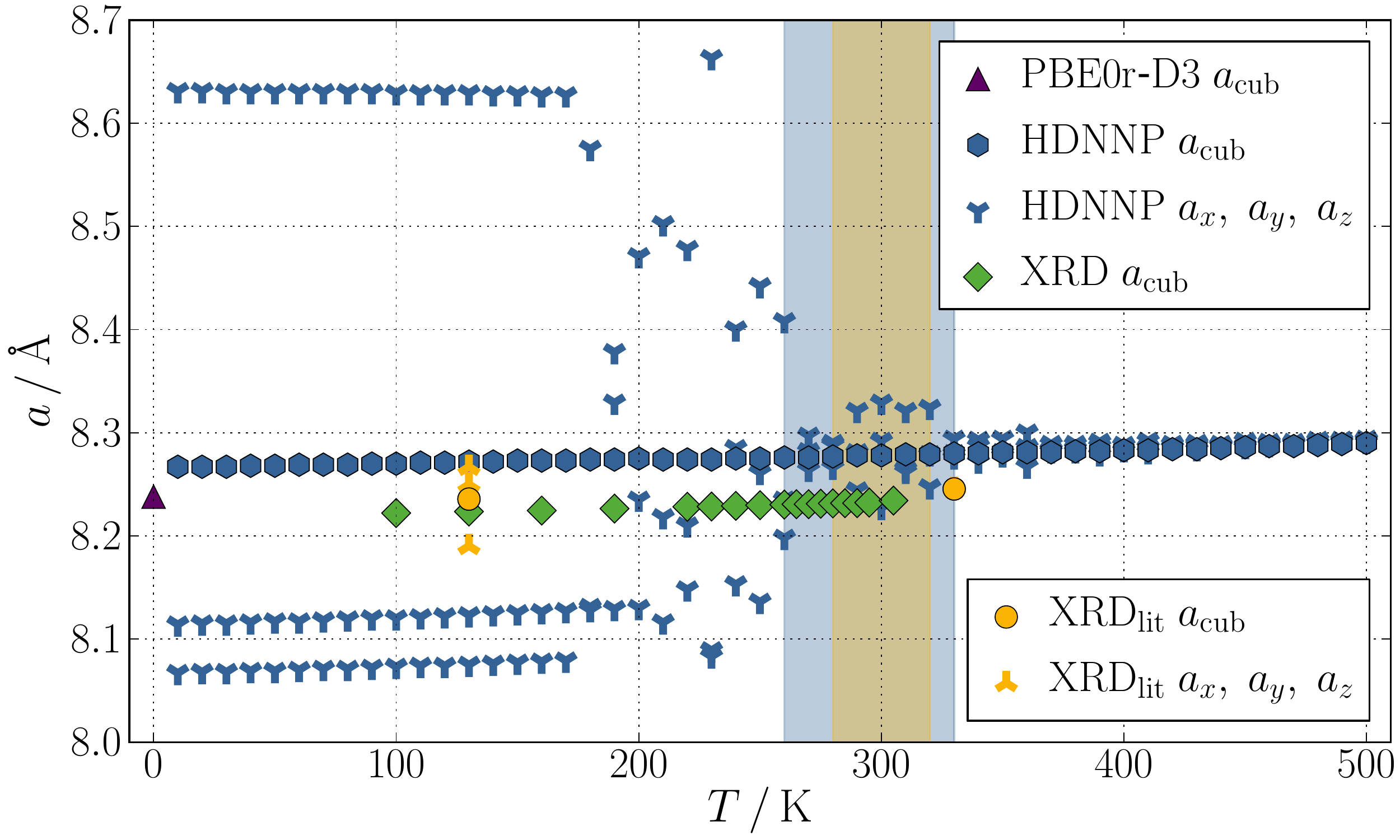}
\caption{Averaged lattice constant $a_\mathrm{cub}$ and individual lattice constants $a_{x,y,z}$ as a function of the temperature $T$ determined by PBE0r-D3 (0\,K), the HDNNP (40\,ns MD for each $T$), and XRD for LiMn$_2$O$_4$. The XRD$_\mathrm{lit}$ data have been taken from Reference \cite{Akimoto2004}. The temperature region of the transition from an orthorhombic to a cubic crystal structure is highlighted in blue (HDNNP) and yellow (XRD$_\mathrm{lit}$).}
\label{fig_T-a_8_111}
\end{figure}

We start with a discussion of the spatially averaged cubic lattice constants $a_\mathrm{cub}$ in Figure \ref{fig_T-a_8_111} before we analyze the phase transition based on the time-averaged results of $a_x$, $a_y$, and $a_z$. Extrapolating the HDNNP results of $a_\mathrm{cub}$ to 0\,K yields 8.266\,\AA{}, which is identical to the value of the HDNNP optimized structure and agrees very well with the optimized PBE0r-D3 value of 8.238\,\AA{}. The HDNNP only slightly overestimates the lattice constant by about 0.3{\%}. Our XRD result for $a_\mathrm{cub}$ at 130\,K agrees within 0.1{\%} with the data from the literature \cite{Akimoto2004}. Moreover, the theoretical and experimental values are very similar. The deviations of PBE0r-D3 to our 0\,K extrapolated XRD data and the 0\,K extrapolated XRD$_\mathrm{lit}$ data are smaller than 0.3{\%} and 0.1{\%}, respectively. The HDNNP predicts a thermal expansion of $4.4\cdot10^{-5}\,\mathrm{\AA}\,\mathrm{K}^{-1}$, which is in excellent agreement with our XRD result of $5.5\cdot10^{-5}\,\mathrm{\AA}\,\mathrm{K}^{-1}$ and the XRD$_\mathrm{lit}$ result of $4.9\cdot10^{-5}\,\mathrm{\AA}\,\mathrm{K}^{-1}$.

Looking at the individual lattice constants $a_x$, $a_y$, and $a_z$ calculated by a time average over 40\,ns $NpT$ simulations using the HDNNP, we observe an orthorhombic structure at low temperatures whose lattice constants are similar to the optimized result at 0 K. As the temperature is increased above about 200\,K the individual lattice constants become more and more similar. Between about 260 and 330\,K the structure finally converts to a cubic cell. Above 330\,K the time averaged lattice constants are virtually equal. In experiment the transition from the orthorhombic to the cubic structure occurs between 280 and 320\,K (yellow region in Figure \ref{fig_T-a_8_111}) \cite{Shimakawa1997, Akimoto2004}.

\begin{figure}[tb!]
\centering
\includegraphics[width=\columnwidth]{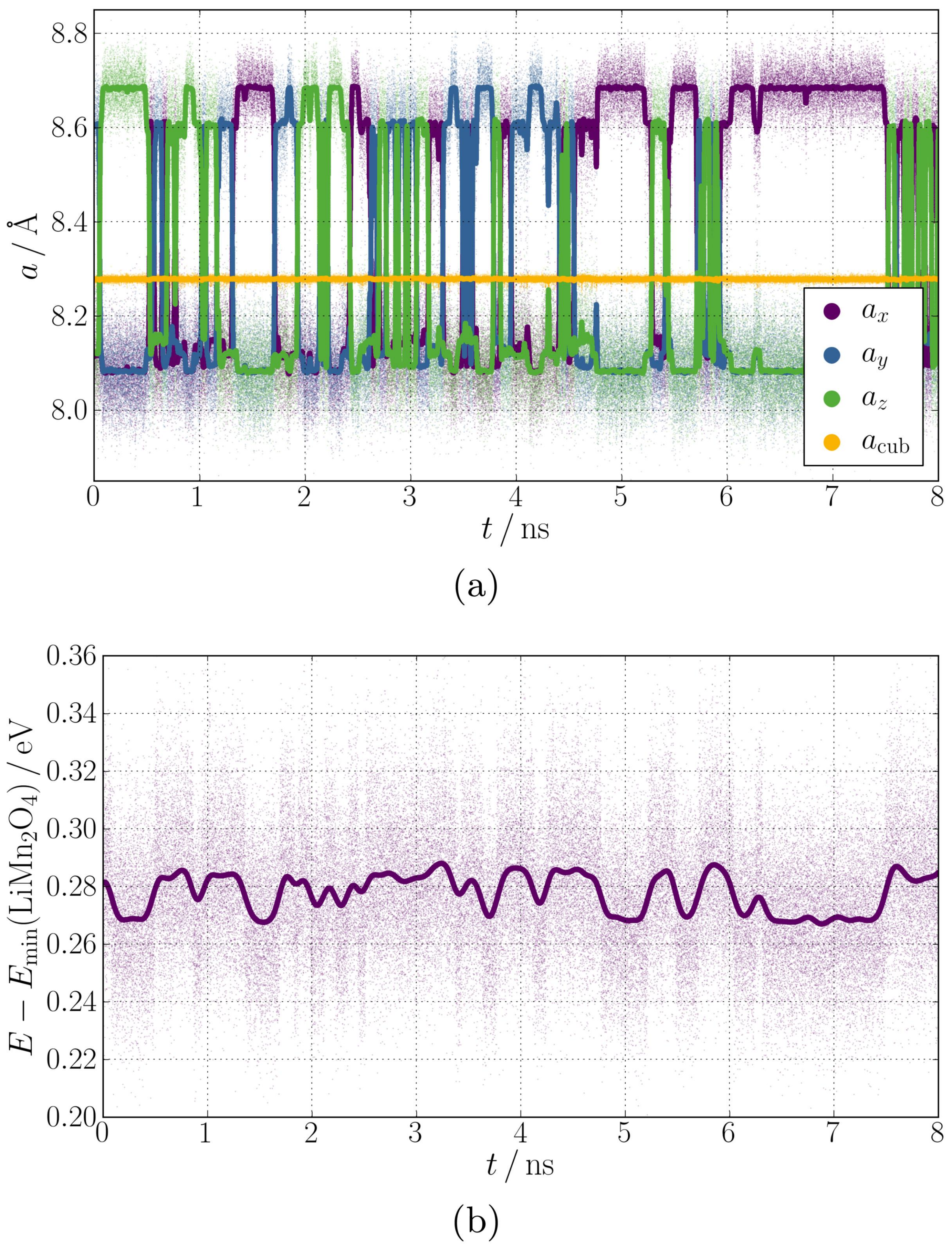}
\caption{(a) Lattice constants $a_{x,y,z}$ and $a_\mathrm{cub}$ and (b) potential energy $E-E_\mathrm{min}(\mathrm{LiMn}_2\mathrm{O}_4)$ per formula unit as a function of simulation time $t$ determined by the HDNNP at 300\,K for LiMn$_2$O$_4$. $E_\mathrm{min}(\mathrm{LiMn}_2\mathrm{O}_4)$ is the potential energy of the HDNNP optimized LiMn$_2$O$_4$ structure. The data were collected each fs of the simulation, then averaged over 100\,fs intervals, and finally smoothed by the Kolmogorov-Zurbenko filter \cite{Zurbenko1986}. The averaged raw data of each 100\,fs interval are shown as scatter plot. The shown data are a representative part of a 40\,ns MD simulation of the unit cell.}
\label{fig_t-aE_8_111_300}
\end{figure}

To investigate the transition in more detail, the lattice constants $a_x$, $a_y$, $a_z$, and $a_\mathrm{cub}$ as well as the potential energy $E$ as a function of the time $t$ are given in Figures \ref{fig_t-aE_8_111_300} (a) and (b) for the LiMn$_2$O$_4$ unit cell at 300\,K determined in the same way as in Figure \ref{fig_t-aE_7_111_150} for $\mathrm{Li}_{0.875}\mathrm{Mn}_2\mathrm{O}_4$ at 150 K. The analysis reveals that three LiMn$_2$O$_4$ structures with slightly different lattice constants are present during the simulation, which result from different Mn$^\mathrm{III}$/Mn$^\mathrm{IV}$ distributions. The crystal structure with highest orthorhombicity, i.e., the largest splitting of the lattice constants (largest lattice constant of about 8.68\,$\mathrm{\AA}$ and two equal shorter ones), is energetically favored. For the second most stable structure the two shorter lattice constants are different, for the third structure they are equal again. Therefore, the first and the third structure are actually tetragonal. The Jahn-Teller distorted Mn$^\mathrm{III}$O$_6$ octahedra are aligned in all three structures. The elongated axis can be oriented in $x$, $y$, or $z$ direction for each structure.

The mean energy difference between the most stable structure and the other two structures, which have similar energies, is about 0.02\,eV per formula unit at 300\,K, i.e., 3\,meV\,atom$^{-1}$. These small energy differences are similar to the RMSE of the HDNNP, and also in the order of the accuracy of the underlying DFT calculations. Therefore, these different structure are effectively almost degenerate in energy. However, the structure effects the dynamics of the lattice constant swaps. Direct swaps of the orientation of the cell elongation of the most stable structure did not happen during the entire time of the 40\,ns simulation. Instead, the system has first to undergo a transformation to one of the two other less stable orthorhombic structures with a smaller splitting before it can change its orientation.

Overall, the $E(t)$ curve of LiMn$_2$O$_4$ at 300\,K exhibits higher fluctuations than the one of Li$_{0.875}$Mn$_2$O$_4$ at 150\,K in Figure \ref{fig_t-aE_7_111_150}. In part, this can be attributed to the lower thermal energy at 150\,K, but the energy differences between the states of different orthorhombicity are smaller and no required intermediate configuration for a swap of orientation is observable for Li$_{0.875}$Mn$_2$O$_4$ at 150 K. It seems that the missing Li atom resulting in one Jahn-Teller distorted Mn$^\mathrm{III}$O$_6$ octahedron less in $\mathrm{Li}_{0.875}\mathrm{Mn}_2\mathrm{O}_4$ reduces the activation energy required for the transition. It would be very interesting to compare the two compositions at the same temperature, but we found that this is very difficult. On the one hand, $\mathrm{Li}_{0.875}\mathrm{Mn}_2\mathrm{O}_4$ is far above the stability region at 300\,K and undergoes very rapid changes of the Jahn-Teller distortions. On the other hand, LiMn$_2$O$_4$ essentially does not show any structural changes at 150\,K.

In general at temperatures below 150\,K, orientational fluctuations do not occur on nanosecond time scales. Therefore, the different crystal orientations do not average with time to a cubic cell. At higher temperatures, swaps become possible and the number of swaps per unit time increases with increasing temperature. Averaged over time this results in more and more similar lattice constants. However, the Mn$^\mathrm{III}$O$_6$ octahedra are still Jahn-Teller distorted and the individual structures of the system remain orthorhombic.

We note that in our simulations the transitions occur very rapidly, but a comparison with experimental time scales has to be made with care. The reason is that the number of atoms in these simulations is much smaller than in macroscopic materials studied in experiment. Figure \ref{fig_t-aE_8_111_300} shows the results obtained for a single unit cell containing eight Mn$^\mathrm{III}$ ions, and a reorientation of the Jahn-Teller distortions of such a system can happen very fast. Therefore, we will investigate larger systems in the next Section \ref{sec_scale}. Results of a $3\times3\times3$ supercell reported below will show that the transition times increase with system size.

\subsection{Role of time and length scales}\label{sec_scale}

At first glance, the predicted transition temperature for LiMn$_2$O$_4$ in Figure \ref{fig_T-a_8_111} fits reasonably well to experiment. However, if we consider that the XRD measurements average over much longer times than 40\,ns, this agreement requires further analysis. Since the orientational fluctuations already start at temperatures lower than 260\,K in the simulations, the HDNNP transition temperature to an averaged cubic structure might be expected to shift to lower temperatures if the averaging is based on longer simulations including more transitions.

Furthermore, the splitting of the orthorhombic lattice constants is considerably smaller in experiment than in theory (see Figure \ref{fig_T-a_8_111}). This might be a consequence of using just a single unit cell in our simulations, which leads to a correlation of the atomic motions and a reduced configuration space of the Mn$^\mathrm{III}$ and Mn$^\mathrm{IV}$ ions. Consequently, the corresponding Jahn-Teller distortions might exhibit an artificial superstructure in the system due to the finite simulation cell, and lead to larger splitting of the lattice constants.

We have seen in the discussion of Figure \ref{fig_t-aE_8_111_300} that only three different minimum structures are present in the simulations of a single unit cell, i.e., the configuration space is limited. XRD experiments at 130\,K reveal a distorted $3\times3\times1$ supercell of the high temperature cubic structure, which should be considered in the simulations. For these reasons, we will now investigate the impact of both the total simulation time as well as of the size of the simulation cell with respect to the transition temperature and the degree of orthorhombicity.

\begin{figure}[tb!]
\centering
\includegraphics[width=\columnwidth]{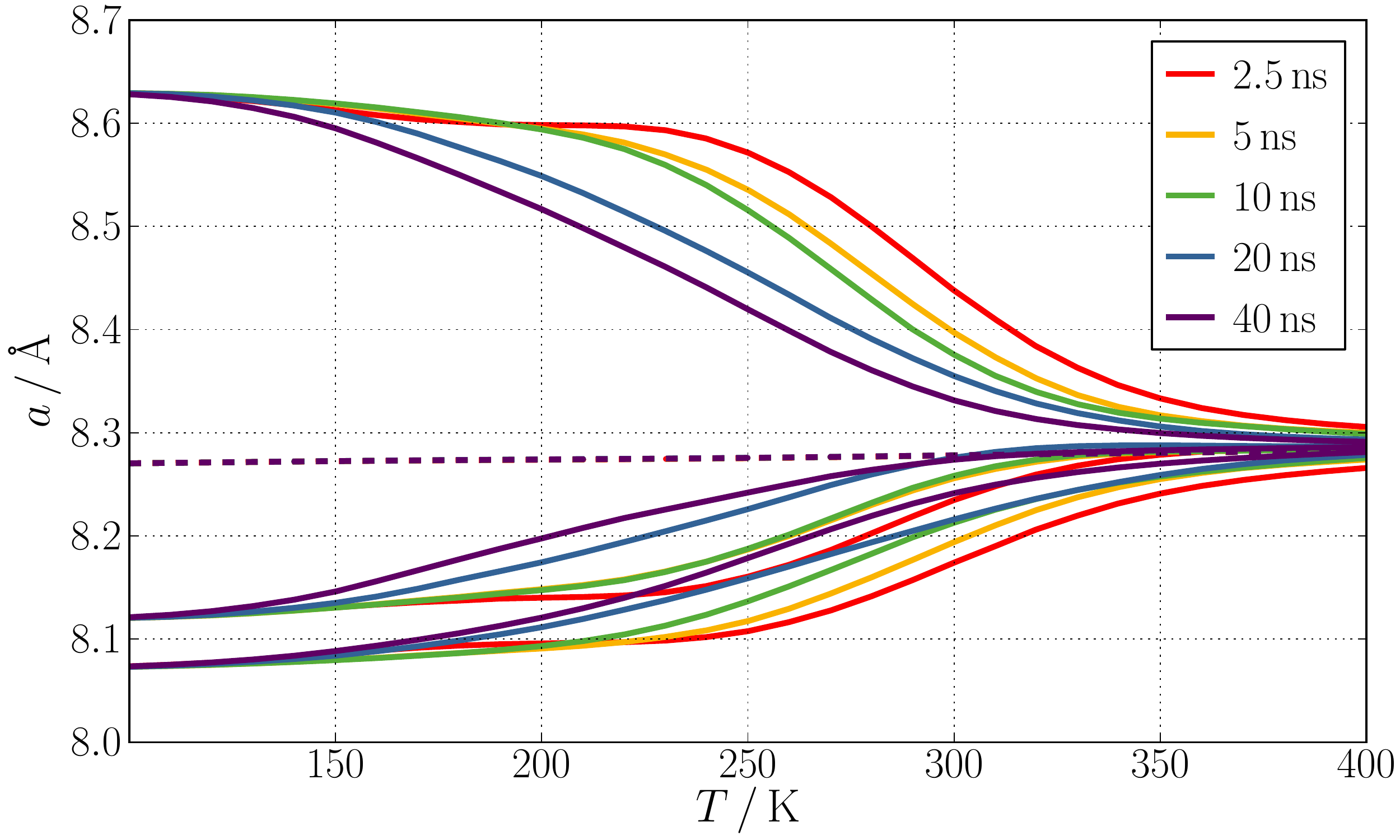}
\caption{Lattice constants $a_{x,y,z}$ (solid lines) and $a_\mathrm{cub}$ (superimposed dashed lines) as a function of the temperature $T$ obtained from fractions of increasing length taken from a 40\,ns $NpT$ simulation of the LiMn$_2$O$_4$ unit cell using the HDNNP. The 40\,ns result includes all other trajectories. The data were smoothed by the Kolmogorov-Zurbenko filter ($k=3$, $m=7$) to highlight the general trend \cite{Zurbenko1986}.}
\label{fig_T-a_111_simulation_time}
\end{figure}

Starting our analysis with the simulation time, we observe that averaging over longer times shifts the transition of the unit cell to lower temperatures (Figure \ref{fig_T-a_111_simulation_time}). As expected, the mean cubic lattice constant $a_\mathrm{cub}$ is almost not affected, but the averages of the individual lattice constants $a_x$, $a_y$, and $a_z$ become more similar with increasing simulation time as more swaps of the elongation direction occur resulting in better statistics.

\begin{figure}[tb!]
\centering
\includegraphics[width=\columnwidth]{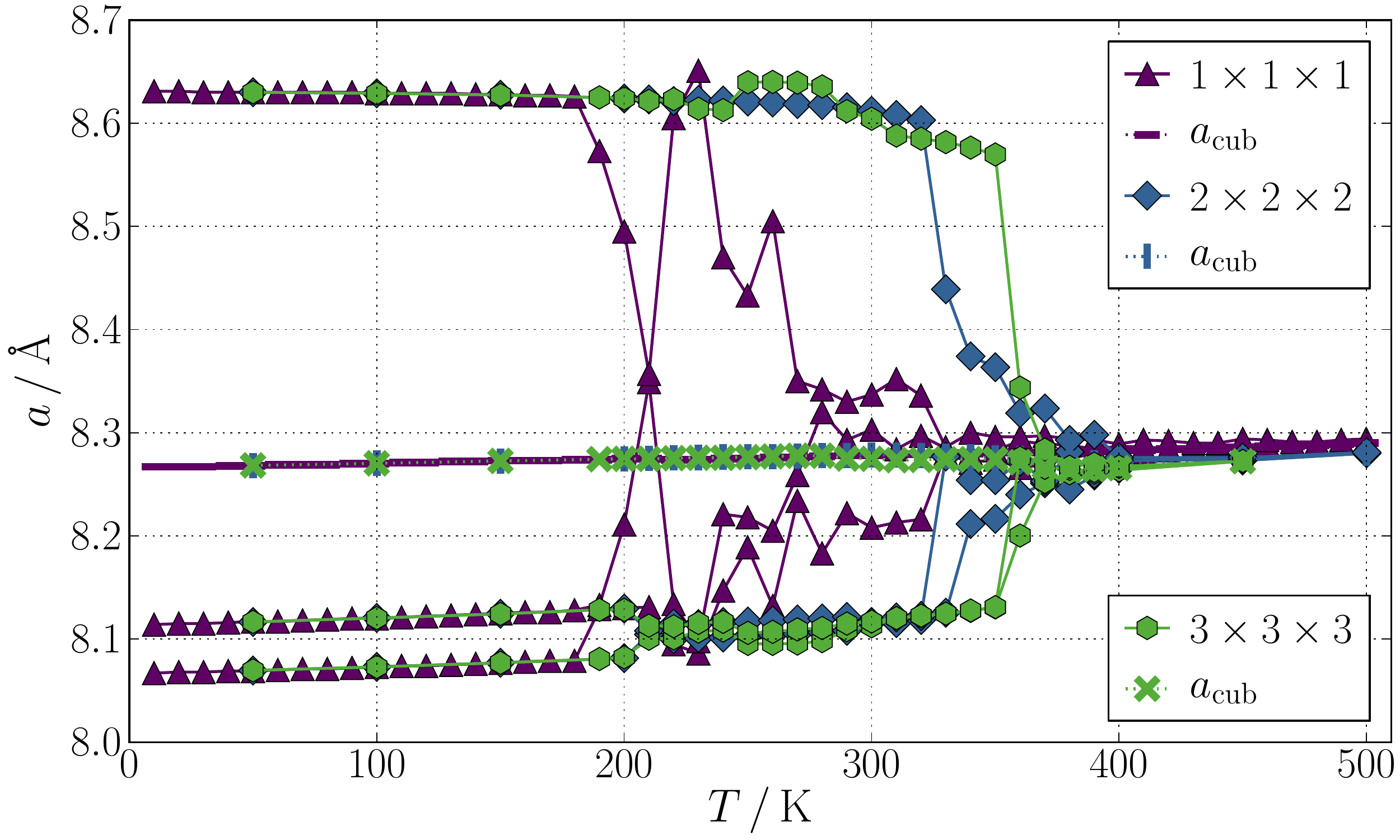}
\caption{Lattice constants $a_{x,y,z}$ and $a_\mathrm{cub}$ as a function of the temperature $T$ obtained from 20\,ns $NpT$ simulations of LiMn$_2$O$_4$ cells of different size using the HDNNP. For facilitating the comparison the lattice constants of the larger supercells have been normalized to the size of a single unit cell.}
\label{fig_T-a_simulation_size}
\end{figure}
 
Phase transitions in finite systems are often shifted in temperature and extend over a wider temperature range than in infinite systems \cite{Challa1986, Binder1987, Wandelt2015}. Finite system means in this context that the simulation cell has a finite size. Due to the periodic boundary conditions we simulate an infinite system but the structure and dynamics of the periodic images are the same as those of the original cell leading to the finite size effects. In a finite system the order parameter changes continuously even for phase transitions of first order. The larger the simulation cell, the closer is the transition temperature to the value in the thermodynamic limit.

Figure \ref{fig_T-a_simulation_size} shows the dependence of the time-averaged lattice constants as a function of the system size for a simulation time of 20\,ns up to a $3\times3\times3$ supercell. According to this plot, the transition occurs in a smaller temperature window for larger systems as expected. Averaging over a longer time will not lead to a large shift of the phase transition temperature anymore because especially for the $3\times3\times3$ supercell the transition is more sharply defined. However, the transition temperature increases with system size, which is surprising because larger systems often have lower barriers due to the increase in available configuration phase space.

\begin{figure}[tb!]
\centering
\includegraphics[width=\columnwidth]{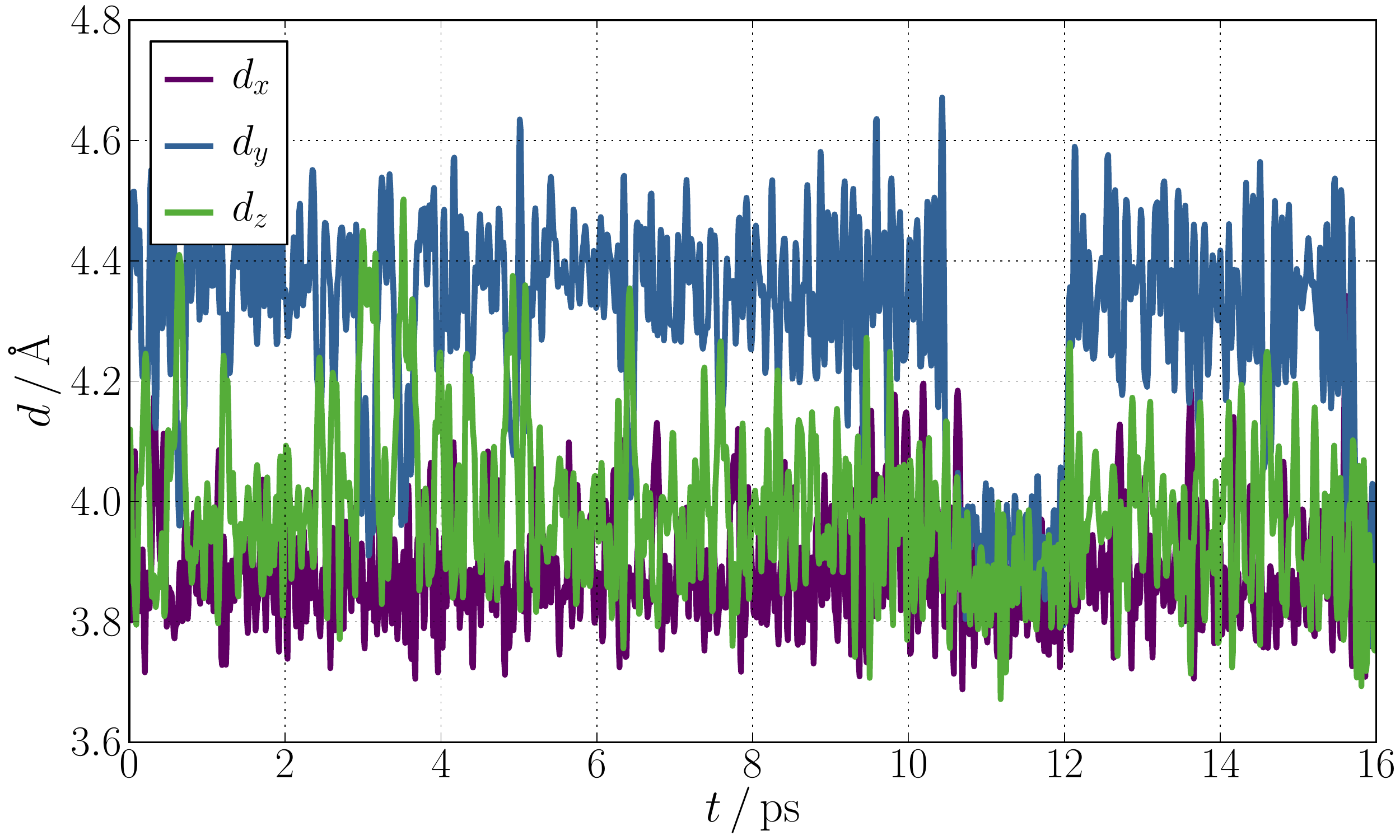}
\caption{Distances $d_{x,y,z}$ of opposite O ions in a MnO$_6$ octahedron as a function of the time $t$ obtained from a HDNNP $NpT$ simulation of a $3\times3\times3$ LiMn$_2$O$_4$ supercell at 300\,K.}
\label{fig_t-d_8_333_300}
\end{figure}

To understand this finding we have to look in more detail at the underlying atomic motions. Therefore, we plotted the distances of opposite O ions $d_x$, $d_y$, and $d_z$ in a typical MnO$_6$ octahedron of a LiMn$_2$O$_4$ $3\times3\times3$ supercell at 300\,K as a function of the time. The lattice constants of a LiMn$_2$O$_4$ $3\times3\times3$ supercell at 300\,K do not swap their orientation in contrast to a single unit cell. However, Figure \ref{fig_t-d_8_333_300} reveals that the Jahn-Teller elongation of an individual Mn$^\mathrm{III}$O$_6$ octahedron is able to change its direction. We can see that for short time intervals the dominant orientation of the largest $d$ value along the $y$ direction of the investigated octahedron changes to the $z$ direction but it always rapidly returns to its initial orientation because the fully aligned configuration is the minimum energy structure. This fluctuation is also a reason for the already slowly decreasing average orthorhombicity of the $3\times3\times3$ supercell above about 290\,K (Figure \ref{fig_T-a_simulation_size}).

Furthermore, in Figure \ref{fig_t-d_8_333_300} it can be observed that the Jahn-Teller distorted Mn$^\mathrm{III}$O$_6$ octahedron with one large and two small $d$ values changes to an undistorted octahedron with three very similar and small $d$ values indicating Mn$^\mathrm{IV}$. We note that the HDNNP does not know explicitly about the electrons. However, as the HDNNP is trained to represent the DFT potential energy surface with high accuracy, which contains all the information about the energetic consequences of the Jahn-Teller distortions as well as e$_\mathrm{g}$ electron hopping, this information is also implicitly included in the HDNNP. Consequently, the MD simulations driven by the HDNNP provide results that would also be obtained by DFT directly, including geometric processes corresponding to e$_\mathrm{g}$ electron hops. Therefore, we can assign the observed geometric changes to these electronic effects, i.e., the Mn ion changes between the Mn$^\mathrm{III}$ and Mn$^\mathrm{IV}$ states due to e$_\mathrm{g}$ electron hops between the Mn sites as the Jahn-Teller distortion appears and disappears. This phenomenon already starts at about 200\,K. 

At low temperatures, no electron hopping and no dynamic Jahn-Teller effect are found, and the Jahn-Teller distorted octahedra are aligned in the simulations. On the other hand, at high temperatures the electron hopping and dynamic Jahn-Teller effect are very fast. This leads to disorder in the Mn$^\mathrm{III}$/Mn$^\mathrm{IV}$ distribution and in the orientation of the Jahn-Teller distortions resulting in an on average cubic structure. Metastable structures exist in $3\times3\times3$ supercells in which the JT distortions are not fully aligned. This is caused by the increased configuration space of the $3\times3\times3$ supercell compared to the single unit cell.

Using this information about the underlying atomic processes we will now develop a simple stochastic model to describe the temperature dependence of the lattice constants swaps as a function of the simulation cell size. 
In our model we only consider the Jahn-Teller distorted Mn$^\mathrm{III}$O$_6$ octahedra. These are elongated in $x$, $y$, or $z$ directions. In the initial state, all octahedra are aligned in the same global direction, for example, the $z$ direction. During a time step in the model, every octahedron can change its orientation to the $x$ or $y$ direction or stay in the $z$ direction with given probabilities. After this distortion step, the $x$ or $y$ direction is considered as candidate for the new global orientation of the system depending on in which direction more changes occurred, for example, the $x$ direction. In case of a tie, the direction is selected randomly. 

The probability that a Jahn-Teller distorted octahedron changes its orientation to the candidate direction for the new global orientation at the given time step in the model is $p(T)$. This probability per time step, i.e., the rate constant, is the same for each octahedron and only a function of the temperature $T$. At higher temperatures the frequency of swaps is larger because the energy barrier can be overcome more easily, and consequently also the frequency of swaps to a new specific direction is higher.

If a critical fraction of the Jahn-Teller distorted octahedra changes the orientations to the candidate for the new global orientation in the same time step in the model, the orientation of the entire structure changes, i.e., also the orientations of the other octahedra are reorientated. Here, we set this critical fraction to $\tfrac{3}{8}=37.5\%$ of the $n$ sites, i.e., three out of eight Mn$^\mathrm{III}$O$_6$ sites in case of a single LiMn$_2$O$_4$ unit cell. We chose this threshold because during the MD trajectories of a single LiMn$_2$O$_4$ unit cell analyzed after each picosecond, we only observed structures with at most one or two Jahn-Teller distorted octahedra that do not align in the majority direction. Structures with three or four octahedra differing in their alignment from the majority direction are only present on a femtosecond time scale. A similar maximum fraction has also been found in larger supercells. If the critical fraction of Jahn-Teller distorted octahedra changes orientation, all other octahedra align into this new direction as well because this is the closest minimum configuration. We tested the model also with different critical fractions, which led to the same conclusions, i.e., the threshold dependence is weak.

If the new orientations of the Jahn-Teller distorted octahedra do not lead to a change of the global orientation of the system, the initial state is restored with every octahedron aligned in the original direction. For initial structures with a high degree of alignment of the Jahn-Teller distorted octahedra as in our simulations, the system will usually be driven back to the aligned minimum configuration unless the temperature is much higher than the transition temperature. The time step in the model is completed after this relaxation and the next time step follows.

The approximations of this model lead to a simple binomial distribution because the Jahn-Teller distorted octahedra can either change to a new global orientation or stay in the old orientation. The probability for a swap of the global orientation is then given by
\begin{align}
P_n(p)=\sum_{k=\tfrac{3n}{8}}^{n}\binom{n}{k}p^k(1-p)^{n-k}\ ,
\end{align}
where $k$ equals the number of orientations changed in the candidate direction for the new global orientation at the given time step in the model. For a swap of the global orientation, $k$ has to be larger than or equal to the threshold.

\begin{figure}[tb!]
\centering
\includegraphics[width=\columnwidth]{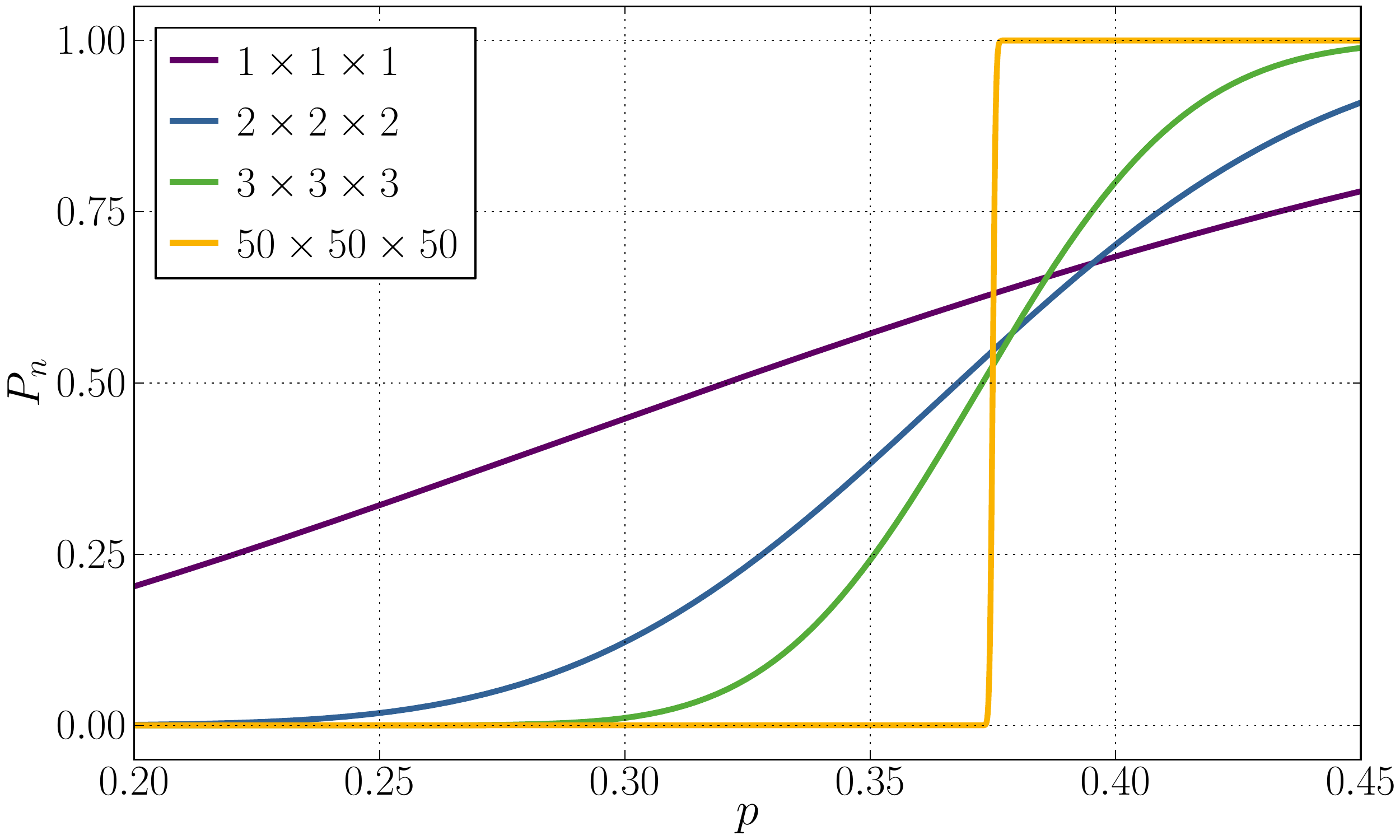}
\caption{Probability for a swap of the global orientation $P_n$ as a function of the probability $p$ for a swap of the orientation of each Mn$^\mathrm{III}$O$_6$ octahedron in the same direction for different cell sizes. The number of Mn$^\mathrm{III}$ sites $n$ in LiMn$_2$O$_4$ is eight times the number of unit cells in the system.}
\label{fig_prob_model}
\end{figure}

Figure \ref{fig_prob_model} shows the dependence of $P_n$ on $p$. Because $p$ is monotonically increasing with temperature we can qualitatively compare the transitions in Figures \ref{fig_T-a_simulation_size} and \ref{fig_prob_model}. Our simple probabilistic model in Figure \ref{fig_prob_model} shows that at low temperatures, successful swaps of the lattice constants are more likely for small cells. For larger cells, swaps are only possible at elevated temperatures equivalent to an increased $p$, i.e., closer to the transition temperature of an infinite system. At high temperatures, the probability for a successful swap of the lattice constants is higher for large cells than for small cells. As a consequence, the temperature window of the transition gets smaller with increasing cell size. For very large cells comparable to experimental nanoparticles, like the $50\times50\times50$ supercell, there is a sharp transition temperature.

The results of the model are all in agreement with the outcome of the simulations shown in Figure \ref{fig_T-a_simulation_size}. Therefore, the observed trend that the swaps of the lattice constants already occur at lower temperatures for smaller cells is just a consequence of the more correlated Mn$^\mathrm{III}$O$_6$ octahedra in the smaller systems, i.e., an overestimation of collective motions. Larger simulation cells would be required for an exhaustive exploration of the full configuration space. In Figure \ref{fig_T-a_simulation_size}, we see that the averaging at higher temperatures of a smaller cell is not as good as for a larger cell which means that the transition window is broader for the small cell. The data of the $3\times3\times3$ supercell simulations are the best and narrowest estimate for the orthorhombic to cubic transition temperature. They yield a transition window from 350 to 380\,K, which is about 60\,K higher than the experimental result.

\subsection{Phase transition}\label{sec_phase_transition}

To investigate the phase transition in more detail and reveal reasons for the overestimation of the transition temperature, we can use the molar heat capacity $C_p$ at constant pressure as a function of temperature, which reveals transitions as peaks for systems of finite size. The heat capacity can be obtained from total energy fluctuations in a $NpT$ MD simulation,
\begin{align}
C_p=\dfrac{N_A}{N_\mathrm{atoms}k_\mathrm{B}T^2N}\sum_{t=0}^{t_N}\Big[E_\mathrm{tot}(t)-\overline{E_\mathrm{tot}}\Big]^2\ ,
\end{align}
with the Avogadro constant $N_A$, the number of atoms in the simulation cell $N_\mathrm{atoms}$, the Boltzmann constant $k_\mathrm{B}$, and the mean temperature of the $NpT$ MD simulation $T$. The fluctuations of the total energy $E_\mathrm{tot}$ are calculated by the variance of the data from all MD time steps, i.e., from $t=0$ to $t_N$ at step $N$. $\overline{E_\mathrm{tot}}$ is the mean of the total energy during a simulation.

For the heat capacity calculations we used the time evolution of the total energy from 20\,ns HDNNP $NpT$ MD simulations at different temperatures of the single unit cell, the $2\times2\times2$ supercell, and the $3\times3\times3$ supercell. We observed a first peak at 220, 210, and 205\,K respectively -- with increasing system size the transition temperature decreases as expected \cite{Challa1986, Binder1987, Wandelt2015}. However, this peak was only present because the initial Mn$^\mathrm{III}$/Mn$^\mathrm{IV}$ distribution in the simulations is only the second lowest minimum configuration which we found in unit cell calculations. If the lowest minimum is employed, this peak is not present. Consequently, the simulations below about 240\,K are dependent on their initial configuration. The peak occurs in the heat capacity calculations because the charge order converts to the lowest minimum once the thermal energy is sufficiently high for this transition. Under experimental conditions, we do not start from this non-equilibrium configuration and will not observe a peak. Therefore, we show here only the results of the equilibrated simulations, i.e., $T\geq240$\,K. However, we can still learn from this first peak which temperature is required for electron hopping processes because e$_\mathrm{g}$ electron hopping is needed for this change of the order of the system. In Figure \ref{fig_T-a_simulation_size} we observe that the two smaller lattice constants of the $2\times2\times2$ and the $3\times3\times3$ supercell become more similar, which is the consequence of this transition. 

\begin{figure}[tb!]
\centering
\includegraphics[width=\columnwidth]{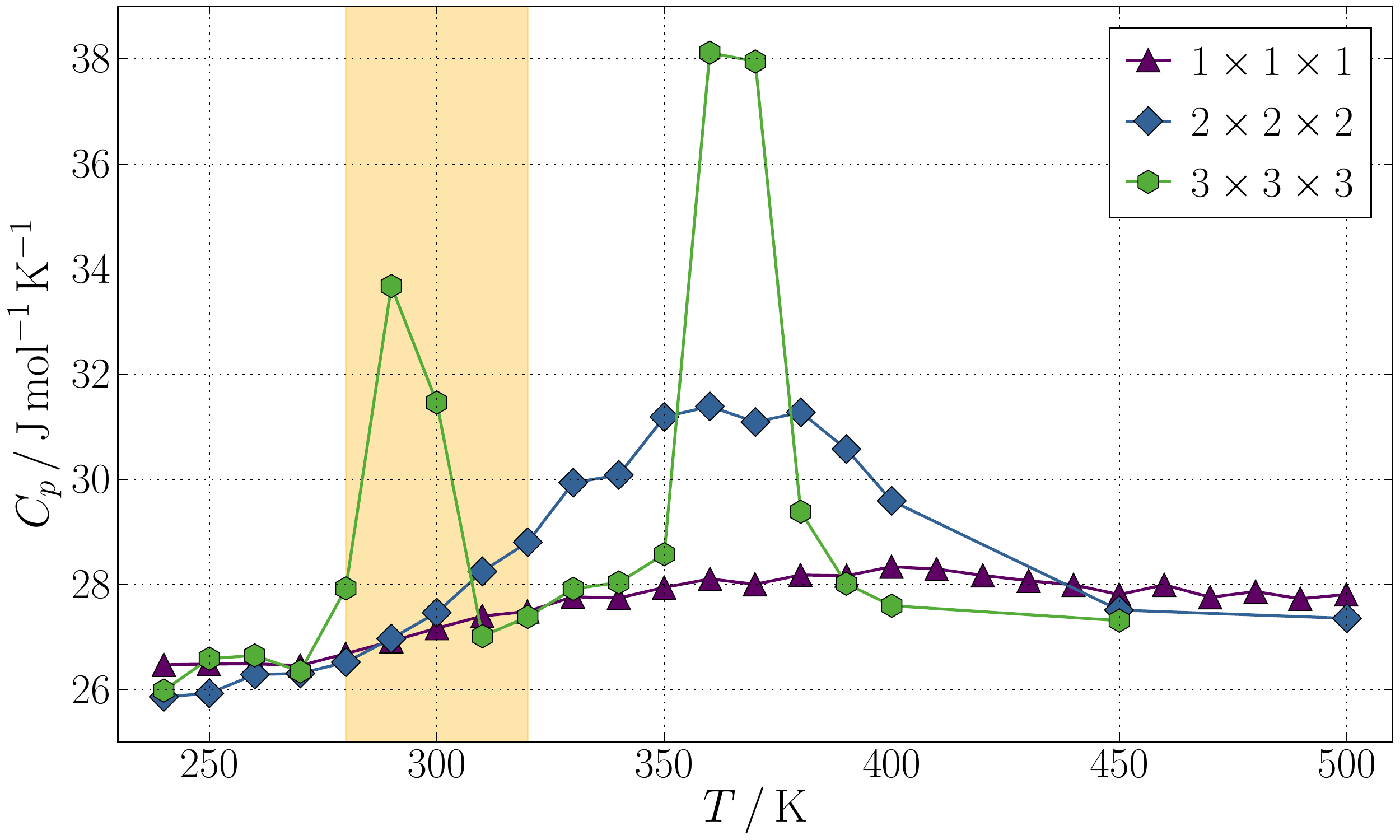}
\caption{Molar heat capacity at constant pressure $C_p$ as a function of the temperature $T$ calculated from 20\,ns HDNNP $NpT$ simulations of an $1\times1\times1$ unit cell and $2\times2\times2$ and $3\times3\times3$ LiMn$_2$O$_4$ supercells at different temperatures. The experimentally determined temperature interval of the orthorhombic to cubic transition using differential scanning calorimetry and single-crystal XRD is highlighted in yellow \cite{Akimoto2004}.}
\label{fig_T-Cp}
\end{figure}

The heat capacity graph in Figure \ref{fig_T-Cp} shows a peak between about 320 and 400\,K for the $2\times2\times2$ supercell as well as between 350 and 380\,K for the $3\times3\times3$ supercell. This peak is very broad in the case of the single unit cell with a maximum around 400\,K and gets sharper and more pronounced for larger cells as expected \cite{Challa1986, Binder1987, Wandelt2015}. For the $2\times2\times2$ and $3\times3\times3$ supercell it occurs in the same temperature range where the dynamics of the Jahn-Teller distortions leads to a time-averaged cubic structure. The transition can be interpreted analogously to a magnetic transition at the Curie temperature, where the alignment of spins is lost due to thermal energy. In our case, the alignment of the Jahn-Teller distortions is lost.

For the $3\times3\times3$ supercell, an additional peak in the heat capacity at 290\,K is observed in Figure \ref{fig_T-Cp} which matches the beginning of the decrease of the orthorhombicity in Figure \ref{fig_T-a_simulation_size}. The frequency of electron hopping during the simulations increases at this transition temperature. Probably the heat capacity peak at 290\,K can be attributed to a charge ordering transition which will be analysed in more detail in a subsequent study of the electronic dynamics. In the single unit cell and the $2\times2\times2$ supercell the peak at 290\,K is not observable. This might be a consequence of the limited configuration space with a smaller number of minima and too high collectivity.

\begin{figure}[tb!]
\centering
\includegraphics[width=\columnwidth]{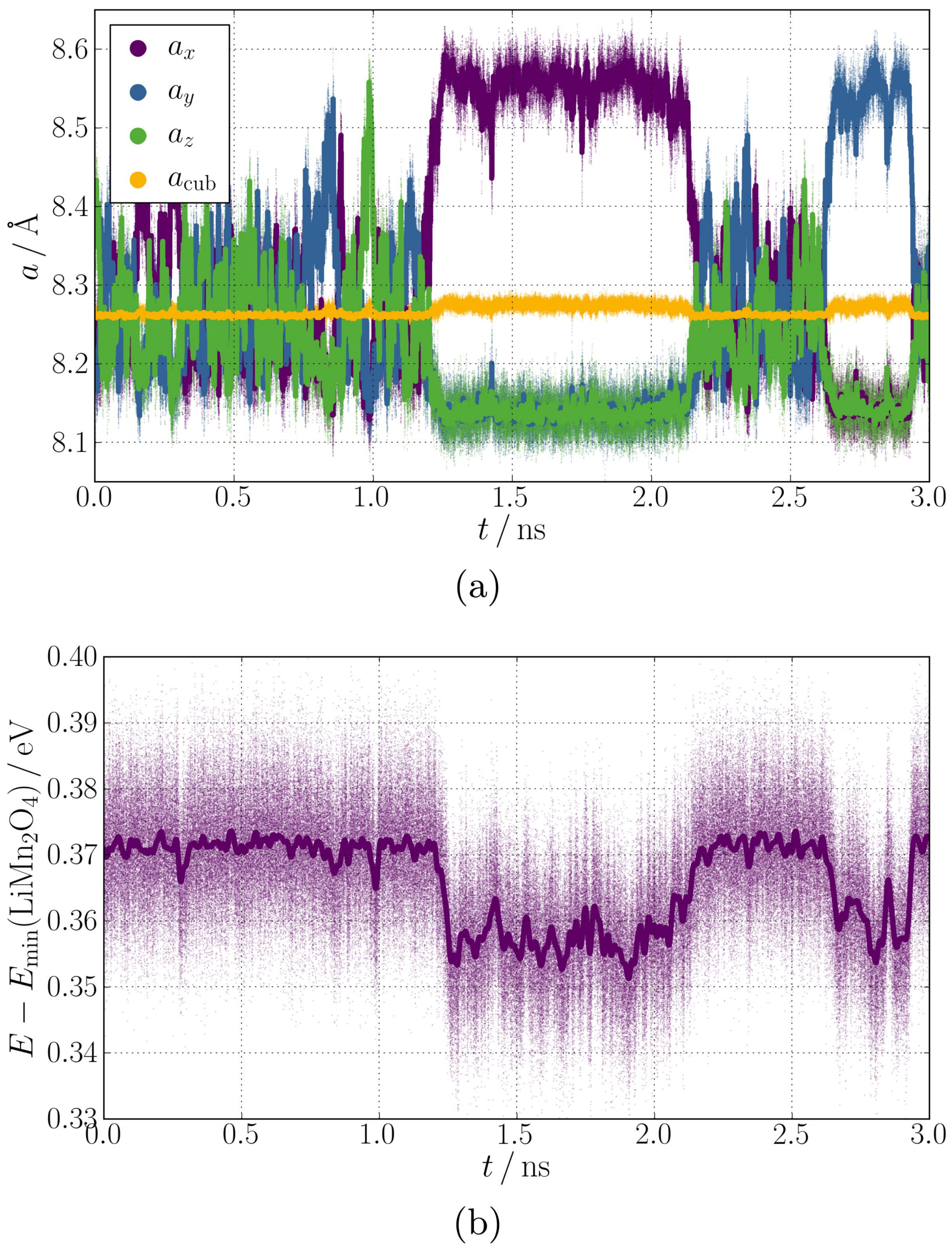}
\caption{(a) Lattice constants $a_{x,y,z}$ and $a_\mathrm{cub}$ and (b) potential energy $E-E_\mathrm{min}(\mathrm{LiMn}_2\mathrm{O}_4)$ per formula unit as a function of the time $t$ determined by the HDNNP at 370\,K for a LiMn$_2$O$_4$ $3\times3\times3$ supercell. $E_\mathrm{min}(\mathrm{LiMn}_2\mathrm{O}_4)$ is the potential energy of the HDNNP optimized LiMn$_2$O$_4$ structure. The data were collected at each fs of the simulation, averaged over 10\,fs, and smoothed by the Kolmogorov-Zurbenko filter \cite{Zurbenko1986}. The averaged data of each 10\,fs are shown as scatter plot. The data represent a typical interval of a 20\,ns simulation.}
\label{fig_t-aE_8_333_370}
\end{figure}

The increase of the configuration space becomes obvious if we compare the time evolution of the lattice constants at the phase transition of the $3\times3\times3$ supercell in Figure \ref{fig_t-aE_8_333_370} with the behavior of a single unit cell in Figure \ref{fig_t-aE_8_111_300}. Figures \ref{fig_t-aE_8_333_370} (a) and (b) show that there is a much larger variety of metastable states present for the $3\times3\times3$ supercell compared to the single unit cell with only three different metastable states and some defect and transitions states occurring in MD simulations at finite temperatures (see Figure \ref{fig_t-aE_8_111_300}). The metastable states of the single unit cell are all orthorhombic crystal structures. The swaps of the orientations lead to a time-averaged cubic structure but the Jahn-Teller distortions are still aligned in the metastable states. Therefore, a spatial average does not lead to a cubic structure in case of the single unit cell. Due to more configurational flexibility in case of the $3\times3\times3$ supercell, the changing orientations of the Jahn-Teller distorted Mn$^\mathrm{III}$O$_6$ octahedra above the transition temperature lead to more or less orthorhombic $3\times3\times3$ supercells depending on the alignment of the individual Jahn-Teller distortions. Metastable configurations exist which are much less orthorhombic than the fully aligned global minimum structure at 0\,K. The geometric mean of these lattice constants is also about 0.015\,$\mathrm{\AA}$ smaller than of the highly orthorhombic structures. Both improve the agreement with experiment. Figure \ref{fig_t-aE_8_333_370} (b) shows that the mean energy of these structures is higher. The sudden changes in volume and energy suggest that the phase transition is of first order, which is in agreement with previous experimental results \cite{Yamada1995}.

The sampling of the $3\times3\times3$ supercell at the phase transition temperature also leads to an answer to the remaining question why our simulations predict a low-temperature orthorhombic structure with larger lattice constant splittings than measured in experiment. For the single unit cell, only configurations with fully aligned Jahn-Teller distortions are metastable. But we observe in Figure \ref{fig_t-aE_8_333_370} (b) that the mean potential energy of the less orthorhombic $3\times3\times3$ supercell structures is only about 2\,meV higher than the fully aligned minimum structure. Consequently, metastable configurations exist for the $3\times3\times3$ supercell energetically close to the aligned minimum configuration, in which the Jahn-Teller distortions are not fully aligned in one direction. Thus for even larger cells, (partially) ordered configurations with different orientations of Jahn-Teller distortions dependent on the Mn sites might exist which are energetically or entropically favored over the aligned configuration. This would result in a less orthorhombic low-temperature structure.

The outcome of previously published high-resolution diffraction experiments of LiMn$_2$O$_4$ \cite{Rodriguez-Carvajal1998, Massarotti1999, Piszora2004, Akimoto2004} is basically a $3\times3\times1$ supercell of the cubic cell with five different Mn sites. Three are attributed to be Mn$^\mathrm{III}$ sites and two are Mn$^\mathrm{IV}$ sites. As the ratio of these sites does not obey the 1:1 ratio of Mn$^\mathrm{III}$ and Mn$^\mathrm{IV}$ ions, eight Mn$^\mathrm{IV}$ ions in a $3\times3\times1$ supercell are placed on Mn$^\mathrm{III}$ sites \cite{Rodriguez-Carvajal1998}. Each of the Mn$^\mathrm{III}$ sites exhibits a Jahn-Teller distortion in a different direction \cite{Piszora2004}. The crystal structure is orthorhombic because the ratio of the Mn$^\mathrm{III}$ sites is 1:2:2 and the eight remaining Mn$^\mathrm{IV}$ ions cannot be equally distributed over these sites. Therefore, the low-temperature orthorhombic structure is a partially ordered configuration. As a consequence, a single simulation cell, which fully represents this experimentally determined complex charge and Jahn-Teller order, has to be much larger than the $3\times3\times1$ supercell. Otherwise the positions of the Mn$^\mathrm{IV}$ ions on the Mn$^\mathrm{III}$ sites exhibit an artificial superstructure, which might lead to energy penalties and to a higher orthorhombicity.

The experimental crystal structure cannot be directly recalculated because it includes too many Jahn-Teller distorted MnO$_6$ octahedra, since the averaged bond distances hide the Mn$^\mathrm{IV}$ ions on Mn$^\mathrm{III}$ sites. Instead, we investigated an ensemble of partially ordered configurations. We performed 10\,ns $NpT$ simulations at 150\,K using initial structures obtained from a $NpT$ simulation of the $3\times3\times3$ supercell at 400\,K. These systems are usually trapped in different local minima and cannot relax to a fully aligned minimum structure in contrast to similar simulations employing a single unit cell. The corresponding structures are less orthorhombic because the Jahn-Teller distortions are not fully aligned. We minimized the lowest local minimum structure found in these simulations at 150\,K. We note that there might also be different configurations with an even lower mean potential energy, but due to the large configuration space compared to the minimum structures of the unit cell at low temperatures, the search for the global minimum becomes a very complex task. Therefore, the following results refer to a calculation of a specific configuration. The minimization yields lattice constants of 24.751, 24.806, and 24.345\,$\mathrm{\AA}$. These are in good agreement with the experimental $3\times3\times3$ supercell of LiMn$_2$O$_4$ at 130\,K with lattice constants of 24.750, 24.801, and 24.570\,$\mathrm{\AA}$ \cite{Akimoto2004}. The energy is only 3.9\,meV\,atom$^{-1}$ higher than the aligned minimum configuration. If we perform a 20\,ns $NpT$ simulation at 150\,K starting from this structure, we obtain time averaged lattice constants of 24.778, 24.885, and 24.292\,$\mathrm{\AA}$. The splitting of the lattice constants is still somewhat larger than the experimental data, but clearly in better agreement to experiment than results of the unit cell simulations, which are limited to three stable minimum structures of high orthorhombicity.

\section{Conclusion}

In this first application of a HDNNP to a system containing a transition metal in multiple oxidation states accompanied by Jahn-Teller distortions, we demonstrate that this method is able to significantly narrow the gap between DFT and experimental observations. The HDNNP method captures the complex potential energy surface of Li$_x$Mn$_2$O$_4$, enabling simulations with an accurate description of the Jahn-Teller dynamics. The experimentally measured orthorhombic to cubic transition is reproduced by the HDNNP simulations revealing information about the underlying atomistic dynamics. Studying this transition by electronic structure methods like DFT directly would be prohibitively expensive, while the HDNNP allows to bridge this gap by increasing the time and length scales of molecular dynamics. In this way, the transition can now be observed in the simulations, although the studied atomistic processes are rare close to the transition temperature.

Excellent agreement between theory and experiment is found for further properties, such as the lattice constants of both crystal structures which deviate by less than 1{\%} from experiment, and for lattice expansion with increasing Li content as well as temperature. Moreover, a series of other properties such as the lithium diffusion barrier, differences of the electrochemical potential, and phonon frequencies is very similar in experiment and theory. This work proves that the HDNNP method is able to represent a system with multiple oxidation states retaining the accuracy of the underlying first principles method and enabling insights into the atomic dynamics from femto- to nanosecond resolution.

\begin{acknowledgments}
This project was funded by the Deutsche Forschungsgemeinschaft (DFG, German Research Foundation) - 217133147/SFB 1073, projects C03 and C05. We gratefully acknowledge the funding of this project by computing time provided by the Paderborn Center for Parallel Computing (PC$^2$) and by the DFG project INST186/1294-1 FUGG (Project No.\ 405832858). For the support of XRD during in situ delithiation experiments, we thank Damien Saurel and Maria Jauregui from CIC energiGUNE in Vitoria-Gasteiz, Spain and Max Baumung from Universit{\"a}t G{\"o}ttingen, Institut f{\"u}r Materialphysik. Helmut Klein and Heidrun Sowa from Universit{\"a}t G{\"o}ttingen, Kristallographie are acknowledged for the support in measuring and discussing the temperature-dependent XRD data. J.B.\ is grateful for a DFG Heisenberg professorship BE3264/11-2 (Project No.\ 329898176).
\end{acknowledgments}

\bibliography{bibliography}

\end{document}